\documentclass[groupedaddress, nofootinbib]{revtex4-1}[11pt]
\pdfoutput=1

\usepackage{graphicx}
\usepackage {amsmath, amssymb, rotating, bbold,mathrsfs}
\usepackage{hyperref}
\usepackage[usenames,dvipsnames]{color}
\usepackage{slashed}
\usepackage{empheq}
\usepackage{verbatim}

\newcommand{\be}{\begin{equation}}
\newcommand{\ee}{\end{equation}}
\newcommand{\nl}{\nonumber \\}
\newcommand{\x}{\chi}
\newcommand{\order}[1]{\mathcal{O}(#1)}
\newcommand{\keV}{\text{ keV}}
\newcommand{\MeV}{\text{ MeV}}
\newcommand{\GeV}{\text{ GeV}}
\newcommand{\TeV}{\text{ TeV}}

\newcommand{\eV}{\text{ eV}}
\newcommand{\hc}{\mathrm{h.c.}}

\newcommand{\mpl}{m_\mathrm{Pl}}
\newcommand{\Mpl}{m_\mathrm{Pl}}

\newcommand{\mdm}{m_{_\text{DM}}}
\newcommand{\vdm}{v}
\newcommand{\Neff}{N_\text{eff}}

\newcommand{\He}{^4\text{He}}
\newcommand{\p}{\varphi}
\newcommand{\D}{{\rm D}}
\def\he#1{^#1 \mathrm{He}}

\newcommand{\Tg}{T} 
\newcommand{\Tke}{T^{X \, \text{eq}}} 
\newcommand{\Tdec}{T^{X \, \text{dec}}} 
\newcommand{\Tnu}{T^{\nu \, \text{dec}}} 
\newcommand{\Tbbn}{T^\text{BBN}} 
\newcommand{\gstarDM}{g_*^X} 
\newcommand{\gstarnu}{g_*^\nu} 
\newcommand{\xiDMinitial}{\xi_X^0} 
\newcommand{\xiDMinitialFourth}{\xi_X^{0 \, 4}} 
\newcommand{\xiDM}{\xi_X} 
\newcommand{\xinuinitial}{\xi_\nu^{\mathrm{SM}}} 
\newcommand{\xinu}{\xi_\nu} 

\newcommand{\lsig}{\lambda_{\sigma}}
\newcommand{\lsh}{\lambda_{\sigma h}}

\newcommand{\musigp}{{\mu_{\sigma}^{\prime\;2}}}

\newcommand{\asig}{a_{\sigma}}
\newcommand{\vbl}{v_{B-L}}
\newcommand{\kd}{\mathrm{KD}}

\let\Im\relax
\DeclareMathOperator\Im{Im}
\let\Re\relax
\DeclareMathOperator\Re{Re}

\definecolor{paleblue}{rgb}{0.69, 0.93, 0.93}  
\definecolor{darkblue}{rgb}{0,0,0.75}
\definecolor{light-gray}{gray}{0.9}

\makeatletter
\def\@seccntformat#1{\csname the#1\endcsname.\quad}

\def\lsim{\mathrel{\raise.3ex\hbox{$<$\kern-.75em\lower1ex\hbox{$\sim$}}}}
\def\gsim{\mathrel{\raise.3ex\hbox{$>$\kern-.75em\lower1ex\hbox{$\sim$}}}}

\begin{document}

\hspace{11.5cm} \parbox{5cm}{SLAC-PUB-17278}~\\
\vspace{1cm}

\title{A Thermal Neutrino Portal to Sub-MeV Dark Matter}

\author{Asher Berlin}

\author{Nikita Blinov}
\affiliation{SLAC National Accelerator Laboratory, 2575 Sand Hill Road, Menlo Park, CA, 94025, USA}

\date{\today}

\begin{abstract}
\vskip 3pt \noindent 
Thermal relics lighter than an MeV contribute to the energy density of the universe 
at the time of nucleosynthesis and recombination. Constraints on 
extra radiation degrees of freedom typically exclude even the simplest of such dark sectors. 
We explore the possibility that 
a sub-MeV dark sector entered equilibrium with the Standard Model after neutrino-photon decoupling, which significantly 
weakens these constraints and naturally arises in the context of neutrino mass generation 
through the spontaneous breaking of lepton number. 
Acquiring an adequate dark matter abundance independently motivates
  the MeV-scale in these models through the coincidence of gravitational, matter-radiation
  equality, and neutrino mass scales, $(m_\text{Pl} / T^\text{MRE})^{1/4} \,
  m_\nu \sim \text{MeV}$. 
This class of scenarios will be decisively tested by future measurements
of the cosmic microwave background and matter structure of the universe.
While the dark sector dominantly interacts with Standard Model neutrinos, large couplings to  
nucleons are possible in principle, 
leading to observable signals at proposed low-threshold direct detection experiments.
\end{abstract}

\maketitle

\setcounter{page}{0}
\thispagestyle{empty}
\tableofcontents
\newpage

\section{Introduction}
\label{sec:intro}

The mass of dark matter (DM) is relatively unconstrained. Demanding that its de
Broglie wavelength is smaller than the typical size of Dwarf galaxies requires
$\mdm \gtrsim 10^{-22} \eV$, while microlensing searches for massive composite
objects imply that $\mdm \lesssim 10^{58}
\GeV$~\cite{Allsman:2000kg,Wyrzykowski:2011tr,Tisserand:2006zx}. However, if DM
acquired its abundance through thermal contact with the Standard Model (SM)
bath, the  viable mass range is significantly reduced and a much sharper 
picture emerges. For concreteness,
we define ``thermal dark matter" in this manner:

\vspace{5mm}
\noindent \emph{{\bf thermal dark matter}: dark matter that acquired its cosmological abundance after entering thermal equilibrium with the Standard Model bath at temperatures much higher than the freeze-out temperature of number-changing interactions.}
\vspace{5mm}

\noindent The canonical example of this scenario is embodied by the Weakly Interacting Massive Particle (WIMP) paradigm, in which DM is assumed to be in thermal contact with the SM bath while relativistic before chemically (and later kinetically) decoupling from the SM while non-relativistic. For $\mdm \gtrsim \text{keV}$, thermal DM is sufficiently cold such that the free-streaming length in the early universe does not suppress the growth of matter perturbations on scales larger than the observed structures in intergalactic gas~\cite{Viel:2013apy,Baur:2015jsy}. For larger masses, perturbative unitarity requires $\mdm \lesssim 100 \TeV$ under the assumption of a standard thermal cosmological history~\cite{Griest:1989wd}. Thus, the thermal DM paradigm drastically restricts the possible mass range.

Although no theoretical inconsistencies arise for small masses, $\mdm \gtrsim \text{MeV}$ is often quoted as a robust lower bound on the mass of any thermal relic~\cite{Ho:2012ug,Steigman:2013yua,Boehm:2013jpa,Nollett:2013pwa,Nollett:2014lwa,Steigman:2014uqa,Green:2017ybv,Serpico:2004nm}. Such limits are usually derived from indirect measurements of the expansion rate of the universe in the radiation-dominated epoch, which can be parametrized in terms of the effective number of neutrino species, $\Neff$. Sub-MeV thermal DM is relativistic at the time of nucleosynthesis and can modify $\Neff$. 
However, the successful predictions of standard Big Bang nucleosynthesis (BBN) and observations of the cosmic microwave background (CMB) 
constrain $\Neff$ to lie near the SM expectation, $\Neff \simeq 3.046$~\cite{Mangano:2001iu,Mangano:2005cc}. 

As originally pointed out in Refs.~\cite{Chacko:2004cz,Chacko:2003dt,Bartlett:1990qq} and recently studied in the context of light DM in Ref.~\cite{Berlin:2017ftj}, constraints on sub-MeV relics can be alleviated if equilibration between the DM and SM sectors occurs after neutrinos have already decoupled from the photon bath. As we will argue below, this process of \emph{delayed equilibration} is characteristic of thermal DM that is much lighter than a GeV. In this work, we investigate a concrete and predictive model in which this scenario naturally arises for DM thermally coupled to SM neutrinos. There has been resurged interest in models of light thermal DM that interacts with neutrinos~\cite{Bertoni:2014mva,Macias:2015cna,Gonzalez-Macias:2016vxy,Batell:2017rol,Batell:2017cmf,Schmaltz:2017oov}, which has largely been driven by the fact that such interactions constitute a simple mechanism to evade strong constraints from late-time distortions of the CMB~\cite{Ade:2015xua}. 

Although our investigation is warranted solely as a proof of concept for sub-MeV thermal relics, the consideration of such models is timely. Various experimental technologies have recently been proposed for the direct detection of thermal DM down to the keV-scale~\cite{Hochberg:2015pha,Hochberg:2015fth,Schutz:2016tid,Knapen:2016cue}. However, below an MeV, the landscape of cosmologically viable models that will be tested by these experiments is rather unclear and under-explored (see Refs.~\cite{Green:2017ybv,Knapen:2017xzo} for detailed investigations of some simplified models). 
While the most minimal versions of the models examined in this work do not give rise to observable signals at these low-threshold detectors, 
variations upon these scenarios yield detectable rates. We will investigate this in more detail towards the end of this work.
Furthermore, as we will discuss below, our setup will be definitively tested by upcoming cosmological observations, such as CMB-S3/S4 (and to some degree 21-cm) experiments.   

The remainder of this paper is structured as follows. In Sec.~\ref{sec:submev}, we review the standard considerations of sub-MeV thermal relics as studied in previous literature. We then discuss in detail how the standard constraints can be alleviated in a model-independent manner in Sec.~\ref{sec:delequil}. 
In Sec.~\ref{sec:model1}, we introduce a simple concrete model motivated by the observed masses and mixing angles of the SM neutrinos. These models predict DM-neutrino couplings of size $\order{10^{-10}} - \order{10^{-9}}$ and independently motivate thermal DM near the MeV-scale through the coincidence of gravitational, matter-radiation equality, and neutrino mass scales, i.e., $\mdm \sim (m_\text{Pl} / T^\text{MRE})^{1/4} \, m_\nu \sim \text{MeV}$. We then turn to the cosmology and possible modes of detection in Secs.~\ref{sec:cosmo} and \ref{sec:signal}. We briefly summarize our results and conclusions in Sec.~\ref{sec:conclusion}. A more detailed discussion on some aspects of the model is presented in Appendix~\ref{sec:majoron_int}. 

\section{Review of Sub-MeV Thermal Relics}
\label{sec:submev}

In this section, we discuss the physics of light relics 
and their effects on the measurements of primordial 
light element abundances and the CMB. 
For the models considered in this work, the main impact of 
the new degrees of freedom is through their contribution to the 
Hubble expansion rate,
\be
H \simeq \left( \frac{8 \pi}{3} \right)^{1/2} ~ \frac{\rho_\text{rad}^{1/2}}{\mpl}
~,
\ee
where $\mpl \simeq 1.22 \times 10^{19} \GeV$ is the Planck mass and we have assumed that the energy content of the universe is dominated by the radiation component, $\rho_{\rm rad}$.
The radiation energy density includes contributions from SM particles ($\gamma$, $e^\pm$, $\nu$) and the dark sector. 
It is conveniently parametrized by the effective number of neutrino species, $\Neff$, such that 
\be
\label{eq:Neff0}
\rho_\text{rad} \equiv \rho_\gamma \left[ 1 + (7/8) ~ ( \xinuinitial )^4 ~ \Neff(T) \right]
~,
\ee
where $\xinuinitial(T) = T_\nu^{\rm SM}/T_\gamma$ is the neutrino-to-photon temperature ratio in the 
standard cosmology (see Eq.~(\ref{eq:xinuSM}) below).
Thus, $\Neff$  is simply the neutrino and dark sector contribution to the total radiation energy density, normalized to the photon bath.
In contrast to the common definition of $\Neff$ as a late-time quantity (only to be evaluated at the time of recombination), 
$\Neff(T)$ in Eq.~(\ref{eq:Neff0}) parametrizes the expansion rate at temperatures below a few MeV.
$\Neff$ can be modified either by changing the actual number of degrees of freedom in the radiation bath or by 
altering $T_\nu/T_\gamma$. 
The notation for these and other relevant temperature scales is compiled in Table~\ref{Tab:notation} for convenience.

Novel evolution of $\Neff(T)$ can modify the predictions of primordial nucleosynthesis and recombination.
The outcomes of these cosmological epochs have been precisely measured and therefore constrain 
non-standard behavior of $\Neff$. Below, we summarize the effects of varying $\Neff$ on aspects related to BBN and 
the CMB and then review how light dark sectors can run afoul of the resulting constraints.

\subsection{Big Bang Nucleosynthesis}
\label{sec:BBN}

$\Neff$ is constrained by observations of light nuclei abundances, as reviewed in, e.g., Ref.~\cite{Olive:2016xmw}.
The abundances of helium-4, $\he4$, and deuterium, $\D$, are measured with a precision of a few percent and therefore provide the most 
sensitive probes of the expansion rate during the epoch of nucleosynthesis.
We now discuss these elements in turn.

In the early universe, neutrons and protons interconvert through weak processes such as $n\, e^+ \leftrightarrow p \,\bar{\nu}_e$. Once the temperature of the photon bath drops below the neutron-proton mass difference, $\sim \text{MeV}$, the neutron-proton ratio is approximately fixed, $n / p \sim \exp{\left[-(m_n - m_p) / T_{np}\right]}$, where $T_{np} \sim 0.8 \MeV$ is the freeze-out temperature. Most of these neutrons are eventually converted into $\He$ due to its large binding energy per mass (the remainder decays or ends up in deuterium or heavier nuclei). Hence, the $\He$ mass fraction can be estimated by a simple counting argument, $Y_p \simeq 2 \, (n/p)\, / \, ( 1 + n/p) \sim 1/4$. Helium-4 is also produced in stars, but its primordial abundance can be observationally inferred, for instance, from measurements of recombination emission lines of ionized gas in low-metallicity dwarf galaxies~\cite{Izotov:2014fga}. 

Primordial nucleosynthesis is the dominant source of deuterium, since it is destroyed in stellar processes. Its abundance provides an additional handle on constraining the expansion rate at temperatures below an MeV. Deuterium also plays a crucial role in the production of $\He$ through such reactions as $\D ~ p \to \gamma ~ \he3$ followed by $^3\text{He} ~ \D \to p ~ \He$. Due to the small values of the deuterium binding energy ($\sim 2 \MeV$) and baryon-to-photon ratio ($\sim 10^{-10}$), the production of light nuclei is delayed until $T \sim 100 \keV$, a phenomenon known as the ``deuterium bottleneck." However, unlike $\He$, once produced, deuterium is easily destroyed. Deuterium burning proceeds through the same reactions as mentioned above 
until $T \sim 50 \keV$. Its primordial abundance can be determined, e.g., through observations of absorption spectra of distant quasars~\cite{Cooke:2013cba}. 

Modifications to $\Neff$ correspond to changes in the Hubble expansion rate.
For $\Neff > 3$, the expansion rate is enhanced, so that weak processes that convert $n\leftrightarrow p$ 
freeze out earlier (at a larger temperature, $T_{np}$). As a result, the neutron-proton ratio, $n/p$, is increased, leading to a larger
primordial $\He$ abundance with $\Delta Y_p \simeq 0.013 ~ \Delta
\Neff$~\cite{Bernstein:1988ad,Cyburt:2015mya}. Deviations in $\Neff$ also modify the predicted
abundance of deuterium. An increased cosmological expansion rate corresponds to
a shorter time-scale for efficient deuterium burning during $T \sim 50 \keV  - 100
\keV$. Hence, for $\Neff > 3$, the predicted deuterium abundance is increased.

If the baryon density is fixed by the observed nuclear abundances, recent detailed studies have determined $\Neff \simeq 2.85 \pm 0.28$~\cite{Cyburt:2015mya} and $\Neff \simeq 2.87 \pm 0.31$~\cite{Foley:2017xex} within $1 \sigma$ during nucleosynthesis. The spread in the inferred value of $\Neff$ is largely determined by the uncertainty in the primordial value of $Y_p$. This can be seen using $\Delta Y_p \simeq 0.004$~\cite{Aver:2015iza} and the parametric relation $\Delta \Neff \simeq  \Delta Y_p \, / \, 0.013 \simeq 0.3$~\cite{Bernstein:1988ad}. The best-fit central value of $\Neff$ additionally depends on the inferred baryon-to-photon ratio, which is largely driven by the observed deuterium abundance. 

\subsection{Cosmic Microwave Background}
\label{sec:CMB}

Observations of the CMB power spectrum are also sensitive to the total radiation energy density at the time of recombination. Detailed analyses of this effect are presented in Refs.~\cite{Bashinsky:2003tk,Hou:2011ec}. We summarize their arguments below. CMB temperature anisotropies on scales smaller than the diffusion length of photons at recombination are exponentially damped, a mechanism known as Silk or diffusion damping~\cite{Silk:1967kq}. 
On the microscopic level, this corresponds to the stochastic process of photons Thomson scattering with free electrons. Hence, the diffusion distance, $r_d$, can be written parametrically as $r_d \sim \sqrt{N} ~ \lambda_\text{mfp}$, where $N$ is the number of scatters, $\lambda_\text{mfp} \sim 1 / n_e \sigma_T$ is the photon mean free path, $n_e$ is the free electron number density, and $\sigma_T$ is the Thomson cross section. 
The diffusion length scale is therefore $r_d \sim \sqrt{1 / H \lambda_\text{mfp}} ~ \lambda_\text{mfp} \sim \sqrt{1 / H n_e \sigma_T}$.  
A larger $\Neff$ (and correspondingly larger $H$) decreases the diffusion damping \emph{distance} scale. As a result, photons travel 
a shorter average distance out of overdensities. 
However, observations of the CMB measure the \emph{angular} scale of diffusion, $\theta_d = r_d / D_A$, where $D_A$ is the angular distance to the surface of last scattering.
$D_A$ is not independently determined, since it depends on the evolution of dark energy from recombination to present. The dependence on $D_A$ can be eliminated by considering the length scale of the sound horizon, $r_s \sim 1 / H$, at the time of recombination. The position of the first acoustic peak in the CMB power spectrum is dictated by the corresponding angular scale, $\theta_s = r_s / D_A$. Hence, the ratio of angular scales $\theta_d / \theta_s = r_d / r_s \sim \sqrt{H / n_e \sigma_T}$ 
is independent of $D_A$. The position of the first peak has been measured to a precision of $5\times 10^{-4}$~\cite{Ade:2015xua}.
Thus, fixing $\theta_s$ to the observed value, the scaling argument above implies that larger $\Neff$ (and hence $H$) 
leads to a larger $\theta_d$, thereby suppressing power in the damping tail of the CMB. Note that the degree of damping at small angular scales is increased for larger $\Neff$, even though the underlying physical diffusion length 
is decreased. This behavior is seen explicitly in full Boltzmann simulations~\cite{Bashinsky:2003tk,Hou:2011ec}. 
The argument above also makes explicit the degeneracy between $\Neff$ and $Y_p$; since $n_e \propto 1 - Y_p$, 
 the effect on $r_d/r_s$ from decreasing $Y_p$ can be compensated by increasing $\Neff$.
This degeneracy is broken by considerations of BBN.

Measurements by the Planck satellite constrain the effective number of neutrino species at the time of last scattering with unprecedented precision, $\Neff \simeq 3.15 \pm 0.23$ at 68\% confidence~\cite{Ade:2015xua}. Although the inclusion of different cosmological datasets modifies this result slightly, we will take this value as a representative benchmark in our analysis. A recent direct measurement of the local Hubble constant, $H_0$, is in tension with the inferred value from Planck data at the level of $\sim 3.4 \sigma$~\cite{Riess:2016jrr}. The inclusion of additional relativistic species at the time of recombination significantly alleviates the tension, favoring $\Delta N_\text{eff} \simeq 0.4$~\cite{Riess:2016jrr,Bernal:2016gxb,Brust:2017nmv,Oldengott:2017fhy}. This is not the case when the ``preliminary" Planck measurements of high-$\ell$ polarization are included, which favor a standard cosmology, but it is possible that this dataset is plagued by low-level systematics~\cite{Ade:2015xua,Aghanim:2015xee}. 

\subsection{Standard Light Relics}
\label{sec:standardrelic}

Neutrinos decouple from the photon bath at a temperature of $\Tnu \sim 2 \MeV$~\cite{Enqvist:1991gx}. A set of sub-MeV hidden sector (HS) particles (collectively denoted as $X$) that is equilibrated with the SM at temperatures below $\Tnu$ can lead to significant deviations in the observed value of $\Neff$. The lightest stable particle of this HS constitutes the DM of the universe. For simplicity, we assume that $X$ couples to the SM neutrinos and that all such particles have a common mass given by $m_X$. We first consider the standard case where $X$ equilibrates with the SM neutrinos \emph{before} the point of neutrino-photon decoupling, as has been investigated in Refs.~\cite{Ho:2012ug,Steigman:2013yua,Boehm:2013jpa,Nollett:2013pwa,Nollett:2014lwa,Steigman:2014uqa,Green:2017ybv,Serpico:2004nm}. The temperature evolution of the neutrino bath is then easily derived from the conservation of comoving entropy density. 

\renewcommand{\arraystretch}{1.5}
\begin{table}
\centering
\begin{tabular}{|c||c|c|}
\hline
\textbf{Notation} & \textbf{Definition} & \textbf{Value} \\
\hline\hline
$T_i$ & temperature of species $i = X, \nu, \gamma$ & $-$ \\ \hline
$T$ & shorthand for the photon temperature ($T_\gamma$) & $-$ \\ \hline
$\xi_i$ &  temperature of species $i$ normalized to the photon temperature & $T_i / T$ \\ \hline
$\Tnu$ & photon temperature at $\nu$-$\gamma$ decoupling & $\sim \order{\text{MeV}}$ \\ \hline
$\Tke$ & photon temperature at $X$-$\nu$ equilibration & $\gg m_X$ (model input) \\ \hline
$\Tdec$ & photon temperature at $X$-$\nu$ chemical decoupling & $\sim m_X$ (model input) \\ \hline
$\Tbbn$ & photon temperature at the end of nucleosynthesis & $\sim \order{10} \keV$ \\ \hline
$T^\text{KD}$ & photon temperature at which $X$ kinetically decouples & $\ll m_X$ (model input) \\ \hline
\end{tabular}
\caption{Notation and various temperature scales discussed throughout this work.}
\label{Tab:notation}
\end{table}

The effective number of relativistic degrees of freedom, $g_*^i$, in each bath ($i = \nu , X , \gamma$) determines the entropy density, $s_i \equiv (2 \pi^2 / 45) \, g_*^i \, T_i^3$, and energy density, $\rho_i \equiv (\pi^2/30) \, g_*^i \, T_i^4$, where $T \equiv T_\gamma$. For three generations of left-handed SM neutrinos,
\be
\gstarnu = (7/8) \times 3 \times 2 = 21 /4
~.
\ee
At temperatures below $\Tnu$, the comoving entropy densities in the $\nu-X$ and photon bath are separately conserved. Using that $s_{\nu + X} \equiv s_\nu + s_X$ and $s_\gamma$ separately scale as $a^{-3}$ ($a$ is the scale factor), one finds
\be
\label{eq:scons1}
\frac{g_*^\nu + \gstarDM}{g_*^\gamma} ~ \xi_\nu^3 = \text{constant}
,
\ee
where
\be
\xi_i \equiv T_i / \Tg
\ee
is the temperature of species $i$ normalized to the photon temperature~\cite{Feng:2008mu}. Treating electron-photon decoupling as instantaneous, we can approximate the number of relativistic degrees of freedom coupled to the photon bath as $g_*^\gamma (T \gtrsim m_e) = 2 + (7/8) \times 4 = 11 / 2$ and $g_*^\gamma (T \lesssim m_e) = 2$. Equating Eq.~(\ref{eq:scons1}) at temperatures above and below $m_e$, and using that $\xi_\nu (T \gtrsim m_e) = 1$, one recovers the standard result 
\be
\xi_\nu (T \lesssim m_e) \simeq \left(\frac{4}{11}\right)^{1/3} \simeq 0.7
~. 
\ee
When $X$ becomes non-relativistic, it heats up the SM neutrinos and negligibly contributes to the entropy density of the $\nu -X$ bath. Again using Eq.~(\ref{eq:scons1}), but for $T_\nu \gtrsim m_X$ and $T_\nu \lesssim m_X$, we find that
\be
\label{eq:xinu1}
\xi_\nu (T_\nu \lesssim m_X) \simeq \left( \frac{4}{11} \right)^{1/3} \left( 1 + \frac{\gstarDM}{g_*^\nu} \right)^{1/3}
~.
\ee
For later convenience, we define $\xinuinitial$ as the value of $\xinu$ assuming a standard cosmology ($\gstarDM = 0$) such that
\be
\label{eq:xinuSM}
\xinuinitial \equiv
\begin{cases}
1 ~,~  \Tg \gtrsim m_e \\
\left( 4/11 \right)^{1/3} ~,~ \Tg \lesssim m_e 
~.
\end{cases}
\ee

Using the above results, the defining expression for $\Neff$ in Eq.~(\ref{eq:Neff0}) can be rewritten as
\be
\label{eq:Neff1}
\Neff (T) \simeq 3 \, \bigg[ \bigg( \frac{\xinu}{\xinuinitial} \bigg)^4 + \Theta(T_X - m_X) ~ \frac{\gstarDM}{\gstarnu} ~ \bigg( \frac{\xiDM}{\xinuinitial} \bigg)^4 \, \bigg]
~.
\ee
In Eq.~(\ref{eq:Neff1}), we have assumed that $X$ decouples instantaneously once its temperature drops below its mass ($T_X \lesssim m_X$), which is encapsulated by the Heaviside step function, $\Theta$~\cite{Feng:2008mu,Berlin:2016gtr}. Note that Eq.~(\ref{eq:Neff1}) reduces to $\Neff \simeq 3$ when $\gstarDM = 0$ and $\xinu = \xinuinitial$. In the SM, neutrino decoupling is not instantaneous, and $e^{\pm}$ annihilations partially heat the neutrino bath, resulting in $\Neff \simeq 3.046$~\cite{Mangano:2001iu,Mangano:2005cc}. 
In Eq.~(\ref{eq:Neff1}), we have approximated $3.046 \simeq 3$. Substituting Eqs.~(\ref{eq:xinu1}) and (\ref{eq:xinuSM}) into Eq.~(\ref{eq:Neff1}), we find that
\be
\label{eq:Neff2a}
\Neff \simeq
\begin{cases}
3 \left( 1 + \gstarDM/g_*^\nu \right) ~,~  T_\nu \gtrsim m_X \\
3 \left( 1 + \gstarDM/g_*^\nu \right)^{4/3} ~,~  T_\nu \lesssim m_X 
~,
\end{cases}
\ee
if $X$ equilibrates with the SM neutrinos at temperatures above $\Tnu$. If $\text{eV} \ll m_X \ll \text{MeV}$, then Eq.~(\ref{eq:Neff2a}) gives $\Neff \gtrsim 3.57$ ($\Neff \gtrsim 3.79$) at the time of nucleosynthesis (recombination) for  $\gstarDM \gtrsim 1$. As discussed in Secs.~\ref{sec:BBN} and \ref{sec:CMB}, this is excluded from considerations of BBN and Planck measurements of the CMB by more than $2 \sigma$. Furthermore, realistic models of light thermal DM often require $g_*^X \gtrsim \text{few}$, leading to even larger deviations in $\Neff$. It is this basic insight that has driven many studies to claim that sub-MeV thermal DM is not cosmologically viable~\cite{Ho:2012ug,Steigman:2013yua,Boehm:2013jpa,Nollett:2013pwa,Nollett:2014lwa,Steigman:2014uqa,Green:2017ybv,Serpico:2004nm}.

\begin{figure}[t]
\hspace{-0.5cm}
\includegraphics[width=0.6\textwidth]{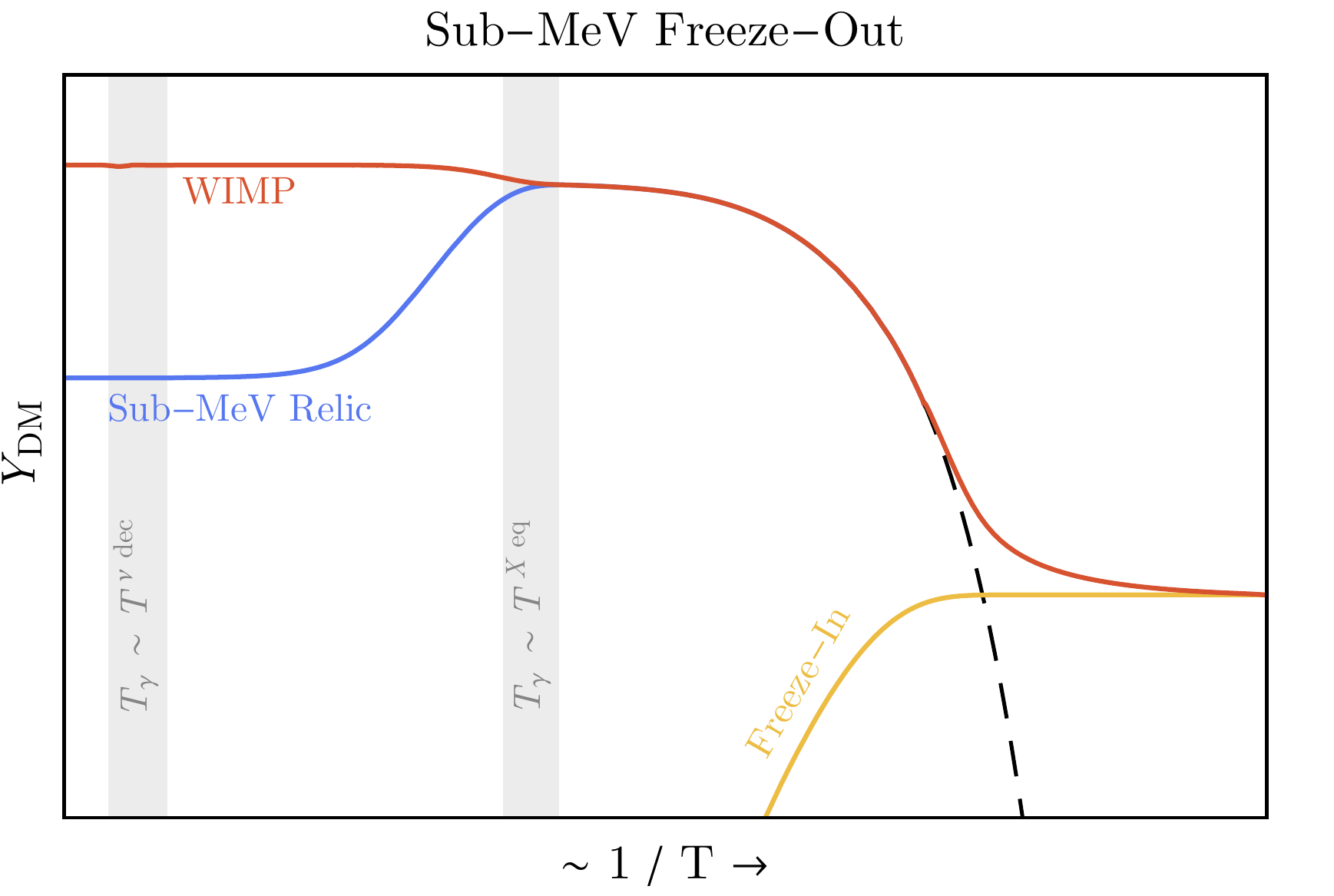} \hspace{-0.5cm}
\caption{The evolution of the dark matter comoving number density ($Y_\text{DM}$) as a function of the photon temperature ($T$). In the standard WIMP framework (red), dark matter is assumed to be in equilibrium with the Standard Model bath long before freeze-out. Dark matter produced through freeze-in (yellow) is assumed to have a negligible abundance at early times and never fully equilibrates with the Standard Model. We propose a scenario (blue) that alleviates strong constraints from measurements of the effective number of neutrino species and is much more akin to the WIMP paradigm, in which an initially cold (compared to the photon bath) population of sub-MeV particles relativistically equilibrates with the Standard Model bath after neutrino-photon decoupling and before freeze-out.  Similar behavior is also expected for standard WIMPs, although the temperature at equilibration ($\Tke$) is typically much larger.}
\label{fig:freezeout}
\end{figure}
%

\section{Delayed Equilibration}
\label{sec:delequil}

\subsection{Temperature Evolution and Effective Number of Neutrino Species}
\label{sec:temp_and_neff_delequil}

In Sec.~\ref{sec:submev}, we noted that a single sub-MeV degree of freedom that is equilibrated with the SM below the temperature of neutrino-photon decoupling, $\Tnu \sim 2 \MeV$, can lead to deviations in $\Neff$ that are in conflict with considerations of BBN and the CMB. In this section, we illustrate that if light relics enter equilibrium with the SM at temperatures below $\Tnu$, then such constraints are significantly relaxed~\cite{Chacko:2004cz,Chacko:2003dt,Bartlett:1990qq,Berlin:2017ftj}.

Let us assume that a similar collection of sub-MeV particles ($X$) equilibrates with the SM neutrino bath while relativistic but \emph{after} neutrino-photon decoupling. The assumption of relativistic equilibration is not strictly necessary, but simplifies the estimates below (see Sec.~\ref{sec:nonrelequil}). As summarized in Table~\ref{Tab:notation}, we define $\Tke \gg m_X$ and $\Tdec \sim m_X$ as the temperature of the photon bath at which $X$ \emph{enters} and \emph{exits} equilibrium with neutrinos, respectively, and $\Tbbn \sim (10- 50) \keV$ as the temperature at which nucleosynthesis has effectively concluded. We will be interested in the case where the HS is initially colder than the SM bath. A schematic representation of the cosmological evolution of the HS comoving number density is shown in Fig.~\ref{fig:freezeout}. Contrary to DM that is produced via freeze-in~\cite{Hall:2009bx}, we assume that the HS is fully relativistic while equilibrating with the SM, analogous to the thermal history of a standard WIMP. 

\begin{figure}[t]
\hspace{-0.5cm}
\includegraphics[width=0.52\textwidth]{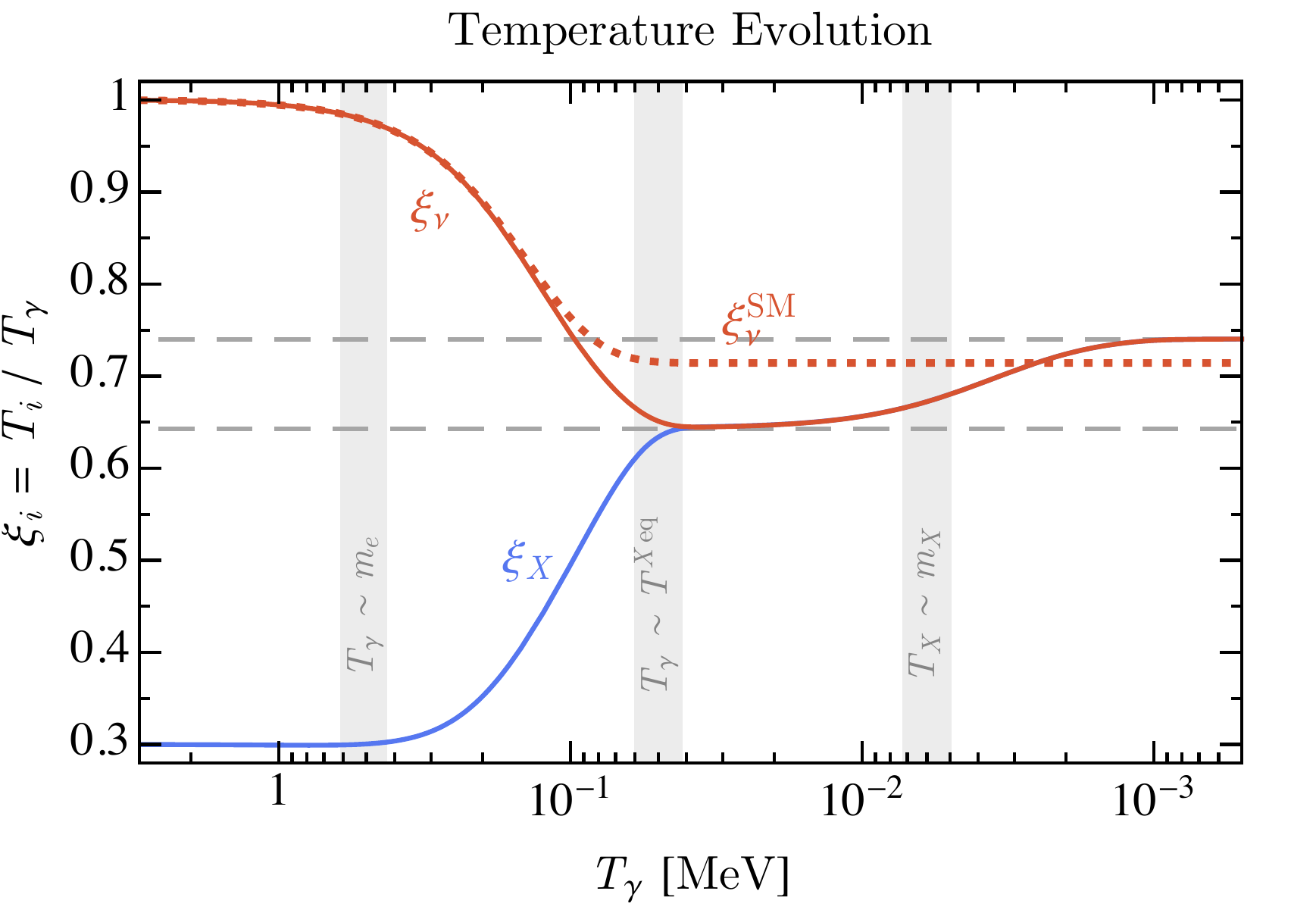} \hspace{-0.5cm}
\includegraphics[width=0.52\textwidth]{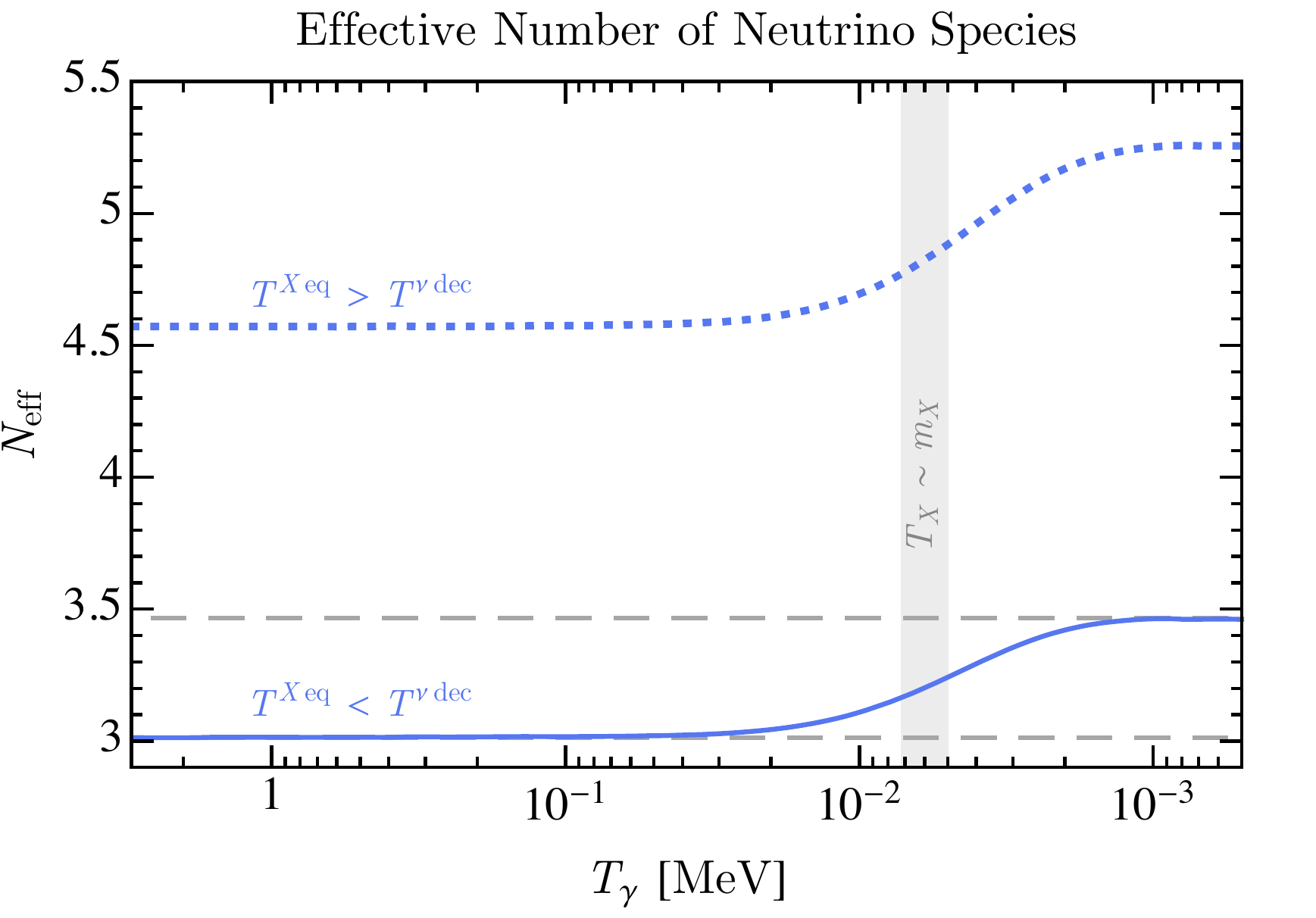} \hspace{-0.5cm}
\caption{(Left) Temperature evolution (normalized to the photon temperature) of the neutrino (red) and dark matter (blue) sectors for an initial temperature ratio of $\xiDMinitial = 0.3$. Compared to standard cosmology, neutrino-dark matter equilibration and decoupling cools and heats the neutrino population relative to its expected value in the Standard Model, respectively. The horizontal gray dashed lines correspond to the approximate analytic estimates of Eqs.~(\ref{eq:xiX1}) and (\ref{eq:xiX2}). (Right) Evolution of the effective number of neutrino species in the case that dark matter equilibrates with neutrinos after (solid blue) or before (dotted blue) neutrino-photon decoupling. The horizontal gray dashed lines correspond to the approximate analytic estimates given in Eqs.~(\ref{eq:Neff3}) and (\ref{eq:Neff4}). For concreteness, we have taken the hidden sector to be made up of a 10 keV Majorana fermion and a 5 keV real scalar. 
}
\label{fig:equil1}
\end{figure}

For concreteness, we assume that $X$ equilibrates with the SM neutrinos after neutrino-photon and electron-photon decoupling, i.e., $\Tke \lesssim \Tnu  \, , \, m_e \sim \text{MeV}$. An example of the temperature evolution of the neutrino and HS baths is shown in Fig.~\ref{fig:equil1}. These results were obtained by numerically solving the Boltzmann equations for the $X$ and $\nu$ energy densities. Analytic approximations will be derived below. If HS-SM equilibration occurs through decays and inverse-decays of a HS species into neutrinos ($X \leftrightarrow \nu \nu$), then the relevant Boltzmann equations are
\begin{align}
\label{eq:Boltzmann1}
\dot{\rho}_X^\text{eq} (T_X) + 3 \, H \, \Big( \rho_X^\text{eq} (T_X) + P_X^\text{eq} (T_X) \Big) &\simeq - \Gamma^\text{dec} \, m_X \, \Big( n_X^\text{eq} (T_X) - n_X^\text{eq} (T_\nu) \Big)
\nl
\dot{\rho}_\nu^\text{eq} (T_\nu) + 4 \, H \, \rho_\nu^\text{eq} (T_\nu) &\simeq + \Gamma^\text{dec} \, m_X \, \Big( n_X^\text{eq} (T_X) - n_X^\text{eq} (T_\nu) \Big)
~,
\end{align}
where $\Gamma^\text{dec}$ is the decay rate for $X \to \nu \nu$, the superscript ``eq'" denotes an equilibrium distribution, and we have been explicit at which temperature the equilibrium number/energy densities should be evaluated. In writing the above equations, we have 
neglected Bose-enhancement and Pauli-blocking factors. Including these effects modifies the collision term by $\order{1}$ factors, but does not significantly change our results. 
Eq.~(\ref{eq:Boltzmann1}) can be solved numerically for the evolution of $T_{X , \nu}$ as a function of the photon temperature, $T$. The time variable can be traded for the photon temperature through the relation~\cite{Venumadhav:2015pla}
\be
\dot{T} = -\, 3 \, H \, \left( \frac{d \rho_\text{tot}}{d T} \right)^{-1} \, \left( \rho_\text{tot} + P_\text{tot} \right)
~,
\ee
where $\rho_\text{tot} \equiv \rho_\gamma + \rho_\nu + \rho_X$ and similarly for the pressure density, $P_\text{tot}$.
In Eq.~(\ref{eq:Boltzmann1}), we have neglected chemical potentials, assuming that the interactions between the HS and neutrino baths enable each species to rapidly track equilibrium distributions dictated by $T_{X , \nu}$. This is a good approximation for the model described below in Sec.~\ref{sec:model1}, since chemical potentials are suppressed by number-changing reactions involving a light spin-0 mediator. In particular, for $\order{1}$ couplings and keV-scale masses in the HS scalar potential of Sec.~\ref{sec:scalar}, $4 \to 2$ self-interactions involving the spin-0 mediator decouple well after DM freeze-out.

In the left panel of Fig.~\ref{fig:equil1}, we show the cosmological evolution of the neutrino and HS temperatures normalized to that of the photon bath as solid red and blue lines, respectively, assuming that the HS consists of a 10 keV Majorana fermion and a 5 keV real scalar. For comparison, we also display the temperature evolution of the neutrino bath in the SM (dotted red), assuming that no new light thermal relics are present ($g_*^X = 0$). The initial HS-SM temperature ratio is fixed to $\xi_X= 0.3$, such that the HS is initially much colder than the SM neutrino and photon populations. Energy conservation then implies that $\nu - X$ equilibration cools (heats) the neutrino ($X$) bath at $T \sim \Tke$. If this occurs after neutrino-photon decoupling, this leaves the photon bath unaffected. Later, when the temperature drops below $m_X$ and the HS decouples, $X$ dumps its entropy back into the neutrinos, reheating them to a temperature slightly above the SM expectation. These two processes, equilibration (neutrino cooling) and decoupling (neutrino heating), have counteracting effects on the neutrino temperature, which lead to a partial cancellation and a significant reduction in modifications to $\Neff$, whose evolution is shown as the solid blue line in the right panel of Fig.~\ref{fig:equil1}. If the HS is initially colder than the SM bath, this cancellation is a direct consequence of  thermodynamics and does not constitute a tuning of the model. For comparison, we also show the temperature evolution of $\Neff$, taking the standard assumption that equilibration occurs before neutrino-photon decoupling (dotted blue), as in Sec.~\ref{sec:standardrelic} and Refs.~\cite{Ho:2012ug,Steigman:2013yua,Boehm:2013jpa,Nollett:2013pwa,Nollett:2014lwa,Steigman:2014uqa,Green:2017ybv,Serpico:2004nm}. If equilibration occurs \emph{after} neutrino-photon decoupling, deviations in $\Neff$ are significantly reduced. 
We now derive analytic approximations for the asymptotic behavior of $\xi_{\nu , X}$ and $\Neff$, which are shown as the horizontal gray dashed lines in Fig.~\ref{fig:equil1}. 

As we will soon see, $\Neff$ is sensitive to the initial value of $\xiDM \equiv T_X / T$ before $X-\nu$ equilibration or electron-photon decoupling, but, similar to DM production via freeze-in, it is insensitive to the particular value of $\xiDM$ as long as $\xiDM \ll 1$~\cite{Hall:2009bx}. We define $\xiDMinitial \equiv \xiDM (T \gtrsim \Tke \, , \, m_e)$ as this initial temperature ratio. As mentioned above, for simplicity, we assume that electron decoupling occurs before DM equilibration. Comoving entropy is conserved as electrons decouple from the photon plasma. Electron annihilations heat photons relative to the neutrino and $X$ baths. Hence, as in Sec.~\ref{sec:standardrelic}, for $\Tke \lesssim T \lesssim m_e$, we have
\be
\xi_\nu ( \Tke \lesssim T \lesssim m_e) \simeq \left( \frac{4}{11} \right)^{1/3} \quad,\quad \xiDM ( \Tke \lesssim T \lesssim m_e) \simeq \left( \frac{4}{11} \right)^{1/3} \xiDMinitial
~.
\ee
Along with Eq.~(\ref{eq:Neff1}), this implies that $\Neff$ is given by
\be
\label{eq:Neff2}
\Neff (T \gtrsim \Tke ) \simeq 3 \, \left( 1 + \frac{\gstarDM}{\gstarnu} ~ \xiDMinitialFourth \right)
.
\ee
This is the standard result for an uncoupled population of dark radiation. 

If the HS and neutrino baths equilibrate while $X$ and $\nu$ are relativistic, the sum of their comoving energy densities, $\rho_{\nu + X} \, a^4$, is approximately conserved. This can be seen from Eq.~(\ref{eq:Boltzmann1}), which implies that $d \left( \rho_{\nu + X} \, a^4 \right) / d t= \rho_{\nu + X} \, a^4 \, H \left( 1 - 3 \, w \right)$, 
where $w \equiv P_{\nu + X} / \rho_{\nu + X}$. When $T_\nu \gg m_\nu$ and $T_X \gg m_X$, we have $w \simeq 1 / 3$ and $d (\rho_{\nu + X} \, a^4 ) / d t \simeq 0$. Therefore, 
\be
\frac{g_*^\nu \, \xi_\nu^4 + \gstarDM \, \xiDM^4}{\left( g_*^\gamma \right)^{4/3}} = \text{constant}
,
\ee
before and immediately after $X-\nu$ equilibration, where we have used $s_\gamma \propto a^{-3}$. Equating this expression at temperatures above and below $\Tke$, we find
\be
\label{eq:xiX1}
\xi_{\nu X} (\Tdec \lesssim T \lesssim \Tke ) \simeq \left( \frac{4}{11} \right)^{1/3} \left( \frac{g_*^\nu + \gstarDM \, \xiDMinitialFourth}{g_*^\nu  + \gstarDM} \right)^{1/4}
~,
\ee
where $\xi_{\nu X} \equiv \xi_\nu = \xiDM$ is the temperature ratio when $X$ is equilibrated with the SM neutrino bath. Comparing the above expression to the standard result of Eq.~(\ref{eq:xinuSM}), we see that for $\xi_X^0 \ll 1$, $\nu -X$ equilibration significantly lowers the temperature of the neutrino bath, i.e., $\xi_{\nu X} \lesssim \xinuinitial$. Eqs.~(\ref{eq:Neff1}) and (\ref{eq:xiX1}) then imply that
\be
\label{eq:Neff3}
\Neff (\Tdec \lesssim T \lesssim \Tke) \simeq 3 \, \left( 1 + \frac{\gstarDM}{\gstarnu} ~ \xiDMinitialFourth \right)
,
\ee
during $X-\nu$ equilibration and before $X$ becomes non-relativistic. Note that Eq.~(\ref{eq:Neff3}) is identical to the expression of Eq.~(\ref{eq:Neff2}). This is consistent with the fact that $d (\rho_{\nu + X} \, a^4 ) / d t \simeq 0$ and that $\Neff$ is defined in terms of the total radiation energy density.

We use conservation of entropy when $X$ becomes non-relativistic and decouples, since this process occurs in equilibrium. Hence,  
\be 
\frac{g_*^\nu \, \xi_\nu^3 + \gstarDM \, \xiDM^3}{g_*^\gamma}= \text{constant}
,
\ee
just before and after $X$ becomes non-relativistic. Equating this expression above and below $\Tdec \sim m_X$ and using Eqs.~(\ref{eq:xiX1}) and (\ref{eq:Neff1}), we find
\be
\label{eq:xiX2}
\xi_\nu \left(T \lesssim \Tdec \right) \simeq \left( \frac{4}{11} \right)^{1/3} \left( 1 + \frac{\gstarDM}{g_*^\nu} \right)^{1/12} \left( 1 + \frac{\gstarDM}{g_*^\nu} \, \xiDMinitialFourth \right)^{1/4}
~,
\ee
 and
\be
\label{eq:Neff4}
N_\text{eff} \left( T \lesssim \Tdec  \right)  \simeq 3 \left( 1 + \frac{\gstarDM}{g_*^\nu} \right)^{1/3} \left( 1 + \frac{\gstarDM}{g_*^\nu} \, \xiDMinitialFourth \right) 
~.
\ee
Note that in the $\xiDMinitial \ll 1$ limit and taking $\Tdec \sim m_X$, Eqs.~(\ref{eq:Neff2}), (\ref{eq:Neff3}), and (\ref{eq:Neff4}) reduce to 
\be
\Neff (T \gtrsim m_X) \simeq 3
\ee
 and
 \be
 \label{eq:NeffMin}
 \Neff (T \lesssim m_X) \simeq 3 \left( 1 + \gstarDM / g_*^\nu \right)^{1/3} \gtrsim 3.18
 ~,
 \ee
where in the inequality we have imposed $\gstarDM \gtrsim 1$ for any light HS. 

Compared to the standard result of Eq.~(\ref{eq:Neff2a}), the deviation in $\Neff$ away from its SM expectation is significantly reduced in Eq.~(\ref{eq:Neff4}) for $\xiDMinitial \ll 1$. As mentioned previously, if $\Tke \lesssim \Tnu$, then $\nu - X$ equilibration drains the neutrino bath of energy, lowering its temperature compared to that of photons. Later, when $X$ becomes non-relativistic and decouples, it reheats the neutrinos to a temperature close to the SM expectation. These processes have counteracting effects on $\xi_\nu$, such that the neutrino bath is reheated to a smaller degree than if $\Tke \gtrsim \Tnu$. However, as seen from Eq.~(\ref{eq:xiX2}), even for $\xiDMinitial \simeq 0$, there is an irreducible heating of the neutrino bath since equilibration of two initially decoupled gases leads to an overall increase in the comoving entropy of the $\nu-X$ system. In the left (right) panels of Fig.~\ref{fig:equil1}, the horizontal gray dashed lines correspond to the approximate values given by Eqs.~(\ref{eq:xiX1}) and (\ref{eq:xiX2}) (Eqs.~(\ref{eq:Neff3}) and (\ref{eq:Neff4})). The numerical solutions are in good agreement with these approximate expressions, which warrants their use in the remainder of this work. 
We also note that a similar cancellation arises when a sub-MeV relic equilibrates directly with the photon bath after neutrino-photon decoupling, but we will not explore such models in this work.

\begin{figure}[t]
\hspace{-0.5cm}
\includegraphics[width=0.6\textwidth]{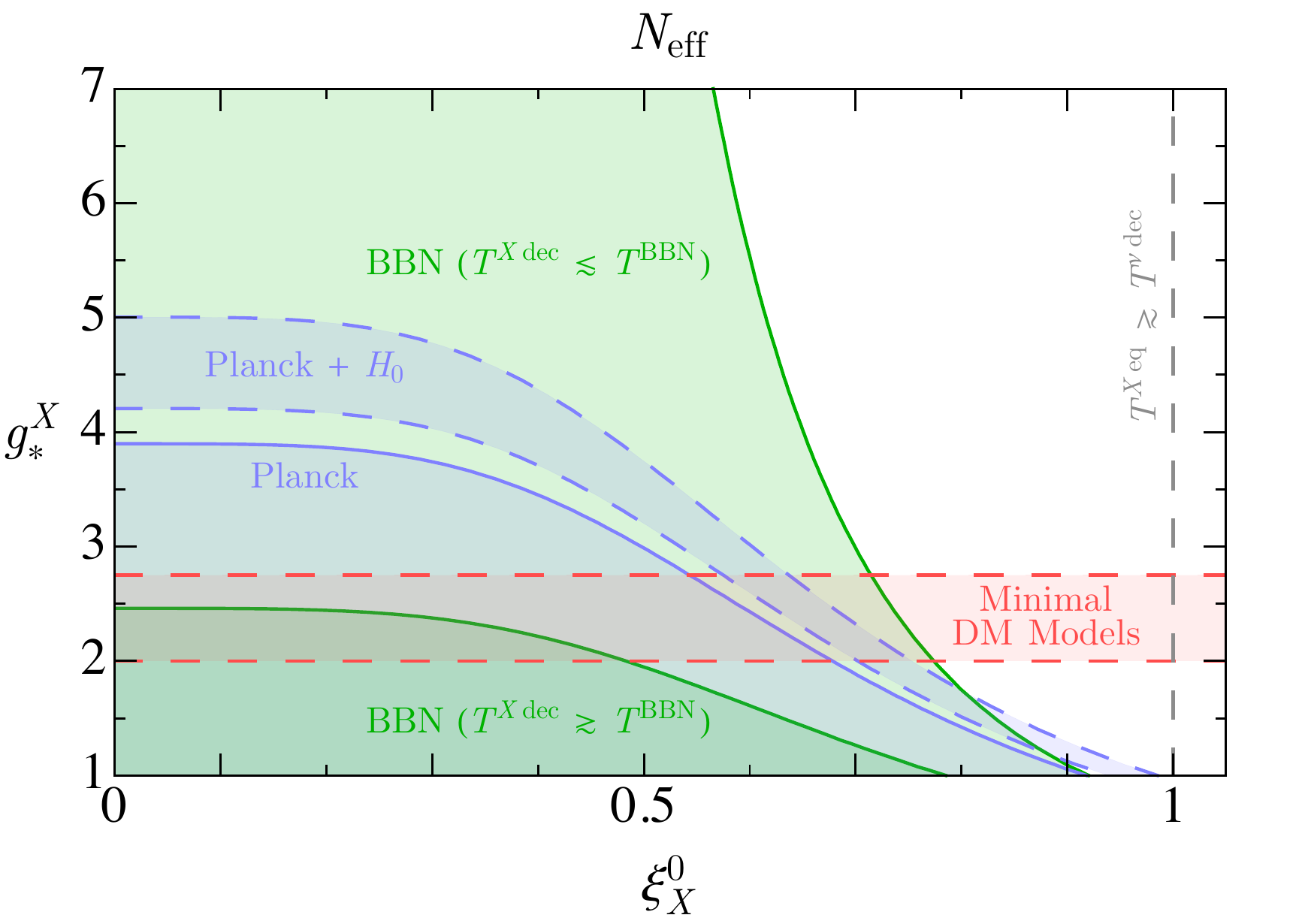} \hspace{-0.5cm}
\caption{Values of $\gstarDM$ (the effective number of sub-MeV  dark sector states that equilibrate with neutrinos) and $\xiDMinitial$ (the initial 
dark sector-to-photon temperature ratio) compatible with the effective number of neutrino species at the time of nucleosynthesis (green) and recombination (blue). Regions compatible with BBN are shown for scenarios in which dark matter decouples from neutrinos before ($\Tdec \gtrsim \Tbbn$) and after ($\Tdec \lesssim \Tbbn$) the end of nucleosynthesis. We also highlight parameter space that alleviates the tension between Planck and local measurements of the Hubble parameter, $H_0$. The representative model space (red) corresponds to a dark sector with a dark matter scalar or Majorana fermion and a scalar mediator. The vertical dashed gray line corresponds to the standard assumption that $X$ equilibrates with neutrinos before neutrino-photon decoupling ($\xiDMinitial \simeq 1$). 
}
\label{fig:Neff_Analytic}
\end{figure}

Equations~(\ref{eq:Neff2}), (\ref{eq:Neff3}), and (\ref{eq:Neff4}) imply that constraints from nucleosynthesis and the CMB can be alleviated if $\Tke \lesssim T^{\nu \text{ dec}}$ and $\xiDMinitial \ll 1$. In Fig.~\ref{fig:Neff_Analytic}, we highlight regions of parameter space in the $\gstarDM - \xiDMinitial$ plane that are compatible with measurements of $\Neff$. If $\Tke \lesssim \Tnu$, then $\xiDMinitial \neq 1$ in general and its value encapsulates the sensitivity of our setup to physics in the ultraviolet. For instance, if $X$ was initially in thermal equilibrium with the SM but decoupled at $T \gtrsim \Lambda_\text{QCD}$ before reentering equilibrium at $T \lesssim \Tnu$, then $\xiDMinitial \sim (10/100)^{1/3} \sim 0.5$. More generally, $\xiDMinitial \neq 1$ arises in theories of asymmetric reheating of the DM and SM sectors~\cite{Adshead:2016xxj}. Throughout this work, we take $\xiDMinitial$ to be a free parameter of the low-energy theory. Note that physics at low-energies is insensitive to this temperature ratio as long as $\xiDMinitial \ll 1$. This is analogous to the level of ultraviolet-sensitivity for DM produced from freeze-in processes, where one typically assumes a negligible initial DM abundance at early times~\cite{Hall:2009bx}.

For $\Tke \lesssim T^{\nu \text{ dec}}$, $\Neff$ transitions from Eq.~(\ref{eq:Neff3}) to Eq.~(\ref{eq:Neff4}) near the decoupling temperature, $\Tdec \sim m_X$.
As a result, limits from nucleosynthesis depend on the ordering of $\Tdec \sim m_X$ and $\Tbbn \sim (10 - 50) \keV$. Regions compatible with BBN are shown in Fig.~\ref{fig:Neff_Analytic} for both of the temperature orderings $\Tdec \lesssim \Tbbn$ and $\Tdec \gtrsim \Tbbn$. For $\Tdec \lesssim \Tbbn$, $\Neff$ is static during BBN and is given only by the expression in Eqs.~(\ref{eq:Neff2}) and (\ref{eq:Neff3}). However, for $\Tdec \gtrsim \Tbbn$, $\Neff$ evolves from the form given in Eqs.~(\ref{eq:Neff2}) and (\ref{eq:Neff3}) to that of Eq.~(\ref{eq:Neff4}) during nucleosynthesis. Detailed studies of BBN, which demand $\Neff \simeq 2.85 \pm 0.28$ within $1 \sigma$, often assume a single fixed value of $\Neff$ throughout the entire formation of light nuclei~\cite{Olive:2016xmw}. However, as we have seen, this is not generally the case for a light HS that equilibrates and decouples from the SM during nucleosynthesis~\cite{Hufnagel:2017dgo}. In deriving a constraint, we demand that $\Neff$ never deviates from the best-fit constant value by more than $2 \sigma$, i.e., $
|\Neff (T)- 2.85| \leq 0.56$ for $T \gtrsim \Tbbn$. We note that this is most likely overly conservative, since for $\xi_X^0 \ll 1$ and values of $\Tdec$ only slightly greater than $\Tbbn$, significant deviations in the expansion rate will only occur at the end of nucleosynthesis.  For instance, this could potentially lead to slight changes in the deuterium or $^7\text{Li}$ abundance without affecting the production of $\He$. It would be interesting to consider the bounds from detailed investigations of BBN, while assuming time-variations of $\Neff$ in this manner. We leave such considerations to future work~\cite{Berlin:2018}. 

Cold DM is necessarily non-relativistic at the time of recombination, i.e., $\text{eV} \ll \Tdec \sim m_X$. To remain consistent with Planck measurements of the CMB within $2 \sigma$, we demand that $|\Neff (T) - 3.15|\lesssim 0.46$ for $T \lesssim m_X$, where we take the form for $\Neff$ given in Eq.~(\ref{eq:Neff4})~\cite{Ade:2015xua}. 
Note that this CMB bound on $\Neff$ assumes standard nucleosynthesis, which is modified in the delayed equilibration scenario, as described above. 
A more realistic approach would be to fit both $Y_p$ and $\Neff$ to the CMB power spectrum. This can significantly expand the allowed 
parameter space due to the $Y_p$-$\Neff$ degeneracy described in Sec.~\ref{sec:CMB}. 
Also shown in Fig.~\ref{fig:Neff_Analytic} are regions of parameter space that alleviate the tension between Planck and local measurements of the Hubble parameter, $H_0$. As a representative favored range, we take $N_\text{eff} \simeq 3.4 \pm 0.05$~\cite{Riess:2016jrr,Bernal:2016gxb,Brust:2017nmv,Oldengott:2017fhy}. Models of light thermal DM require a stable species and a light mediator. We highlight regions of parameter space corresponding to the presence of two real scalars in the HS ($\gstarDM = 2$), or a light Majorana fermion and a real scalar ($\gstarDM = 2.75$). The standard case of $\Tke \gtrsim \Tnu$ corresponds to the limit $\xi_X^0 \simeq 1$, which is in strong tension with measurements of both the CMB and primordial nuclei abundances for $\gstarDM \gtrsim 1$. 

\subsection{General Model-Building}

\begin{figure}[t]
\hspace{-0.5cm}
\includegraphics[width=0.6\textwidth]{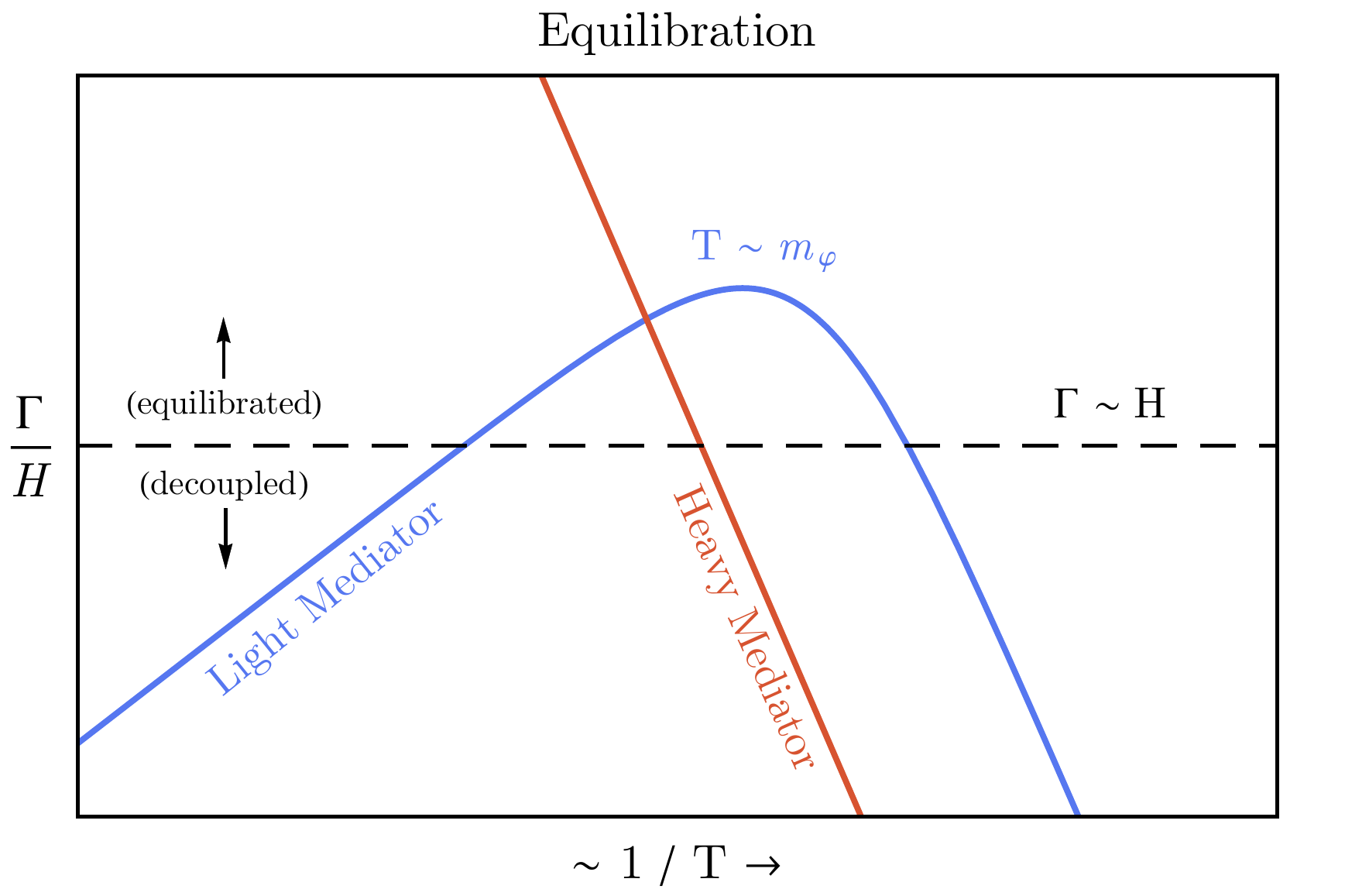} \hspace{-0.5cm}
\caption{$\Gamma / H$ as a function of decreasing temperature for dark matter-Standard Model elastic scattering through the exchange of either a light (blue) or heavy (red) mediator, $\p$. For $\Gamma / H \gtrsim 1$, the hidden sector is in thermal contact with the Standard Model bath. Light mediators generically predict that dark matter enters equilibrium with the Standard Model bath before decoupling.
}
\label{fig:equil}
\end{figure}

We have demonstrated that constraints on sub-MeV thermal relics are weakened when the HS equilibrates with the SM after neutrino-photon decoupling. 
We would like to understand if this naturally occurs in models of light thermal DM.
It has long been appreciated that thermal DM which couples to the SM solely through the electroweak force must be heavier than the GeV-scale. The so-called Lee-Weinberg bound relates the mass of thermal DM to the weak scale ($m_W$), the temperature at matter-radiation equality ($T^\text{MRE} \sim 0.8 \eV$), and the Planck mass ($\mpl$), such that $\mdm \gtrsim m_W^2 / (T^\text{MRE} \, \mpl)^{1/2} \sim \text{GeV}$~\cite{Lee:1977ua}. Equivalently, thermal DM that is lighter than a GeV often requires the presence of new light mediators~\cite{Boehm:2003hm}. It is therefore natural to expect that sub-MeV thermal DM, denoted by $\x$, is accompanied by additional HS mediators, $\p$, that are nearby in mass. 
In this case, there are two processes that can equilibrate the two sectors: 
scattering between HS and SM states, and decays of $\p$ into the SM. As we will show, the temperature dependence of either of these processes generically predicts that a light HS \emph{enters} thermal equilibrium with the SM while relativistic. 
This is illustrated in Fig.~\ref{fig:equil}.
The equilibration point is independent of HS mass scales for scattering, but for decays, it occurs later as HS masses are lowered.
If this proceeds at temperatures below a few MeV, the mechanism described in Sec.~\ref{sec:temp_and_neff_delequil} is realized and modifications to $\Neff$ during nucleosynthesis and recombination are reduced.

At temperatures much greater than $m_\x$ or $m_\p$, we parametrize the rate for scattering and decays/inverse-decays as 
\begin{align}
\Gamma_\text{scatt} &\sim \alpha_\text{eq}^2 \, T ~~ \text{(scattering)}
\nl
 \Gamma_\text{dec} &\sim \alpha_\text{eq} \, m_\p^2 / T ~~ \text{(decays)}
 ~,
 \end{align}
where $\alpha_\text{eq}$ is the effective coupling governing equilibration and the factor of $m_\p / T$ in the decay rate is a time-dilation factor. Comparing either process to the Hubble parameter, $H \sim T^2 / \mpl$, demonstrates that the rate for equilibration overcomes the expansion rate at temperatures below
\begin{align}
\label{eq:Tke}
\Tke &\sim \alpha_\text{eq} \, \mpl ~~ \text{(scattering)}
\nl
\Tke &\sim (\alpha_\text{eq} \, m_\p^2 \, \mpl)^{1/3} ~~ \text{(decays)}
 \end{align}
for scattering and decays, respectively, where $\Tke$ denotes the temperature at which the DM and SM sectors equilibrate. If we parametrize the rate for DM annihilation during freeze-out as $\sigma v \sim \alpha_\text{FO}^2 / m_\x^2$, then $\x$ acquires an abundance in agreement with the observed DM energy density for
\be
m_\x \sim \alpha_\text{FO}  \, (T^\text{MRE} \, \mpl )^{1/2}
~,
\ee
where $\alpha_\text{FO}$ is the effective coupling governing freeze-out. Using this relation in Eq.~(\ref{eq:Tke}) allows us to write $m_\x$ in terms of $\Tke$,
\be
\label{eq:wimp1}
m_\x \sim 
\begin{cases}
(\alpha_\text{FO} / \alpha_\text{eq}) ~ (T^\text{MRE} ~ \Tke)^{1/2}  ~~ \text{(scattering)} \\
(\alpha_\text{FO} / \alpha_\text{eq})^{1/3} ~ (m_\x / m_\p)^{2/3} ~ (\Tke / \mpl)^{1/6} ~ \Tke ~~ \text{(decays)}
~.
\end{cases}
\ee
Equation~(\ref{eq:wimp1}) implies that $\x$ and $\p$ equilibrate with the SM after neutrino-photon decoupling ($\Tke \lesssim \Tnu \sim \text{MeV}$) if 
\be
\label{eq:wimp2}
m_\x \lesssim \text{keV} \times 
\begin{cases}
(\alpha_\text{FO} / \alpha_\text{eq})  ~~ \text{(scattering)} \\
(\alpha_\text{FO} / 10^5 \, \alpha_\text{eq}) ~ (m_\x / m_\p)^{2/3} ~~ \text{(decays)}
~.
\end{cases}
\ee

Bounds on warm DM typically exclude $m_\x \lesssim \text{few} \times \text{keV}$~\cite{Viel:2013apy,Baur:2015jsy}. Therefore, $m_\x \gtrsim \text{keV}$ along with Eq.~(\ref{eq:wimp2}) motivates $\alpha_\text{FO} \gg \alpha_\text{eq}$. This can be accomplished if the processes governing freeze-out are enhanced compared to those governing equilibration. This is a natural hierarchy, for instance, in models of secluded DM~\cite{Pospelov:2007mp}, those involving freeze-out through resonant annihilations~\cite{Griest:1990kh}, or strongly interacting hidden sectors~\cite{Hochberg:2014dra}. Once $\x$ and/or $\p$ become non-relativistic, $\Gamma_\text{scatt}$ and $\Gamma_\text{dec}$ are either suppressed by Boltzmann or $T / m_{\x , \p}$ factors. At this point, the equilibration rate quickly drops below Hubble expansion and the HS decouples from the SM. This behavior can be contrasted with equilibration through the exchange of a heavy mediator, in which case the rate governing equilibration always falls faster in temperature than $H \sim T^2 / \mpl$. 
This is typical of the weak processes that maintain $\nu$-$e$ equilibrium where $\Gamma_{\mathrm{scatt}}\sim G_F^2 \, T^5$.
Schematic examples of these scenarios are shown in Fig.~\ref{fig:equil}. 

The presence of light mediators is strongly motivated for sub-GeV thermal DM. Thermalization through these light mediators generically predicts that DM \emph{enters} equilibrium with the SM while relativistic and before DM freeze-out, as highlighted in Fig.~\ref{fig:equil}. If DM is sufficiently light and there exists a hierarchy between the couplings governing freeze-out and those governing scattering/decays, then the HS equilibrates with the SM after neutrino-photon decoupling, alleviating constraints from measurements of $\Neff$. In Sec.~\ref{sec:model1}, we turn our attention to a concrete model that explicitly realizes this mechanism. However, as an aside, we first briefly comment on scenarios in which the HS instead does not equilibrate with the SM bath until it is semi- or non-relativistic.

\subsection{Non-Relativistic Equilibration}
\label{sec:nonrelequil}

In the previous sections, we focused on a scenario that is closely related to the standard WIMP paradigm:  the HS and SM baths are in equilibrium at temperatures much greater than the DM mass, with chemical decoupling from the SM occurring at temperatures much lower than the DM mass. This is to be contrasted with freeze-in production, in which case DM never fully equilibrates with the SM~\cite{Hall:2009bx}. Although it is not the central focus of this work, an interesting situation may arise between these two extremes, where the HS fully equilibrates with the SM while the DM is semi- or non-relativistic, but before freeze-out of number-changing interactions. We briefly comment on this possibility here.

\begin{figure}[t]
\hspace{-0.5cm}
\includegraphics[width=0.6\textwidth]{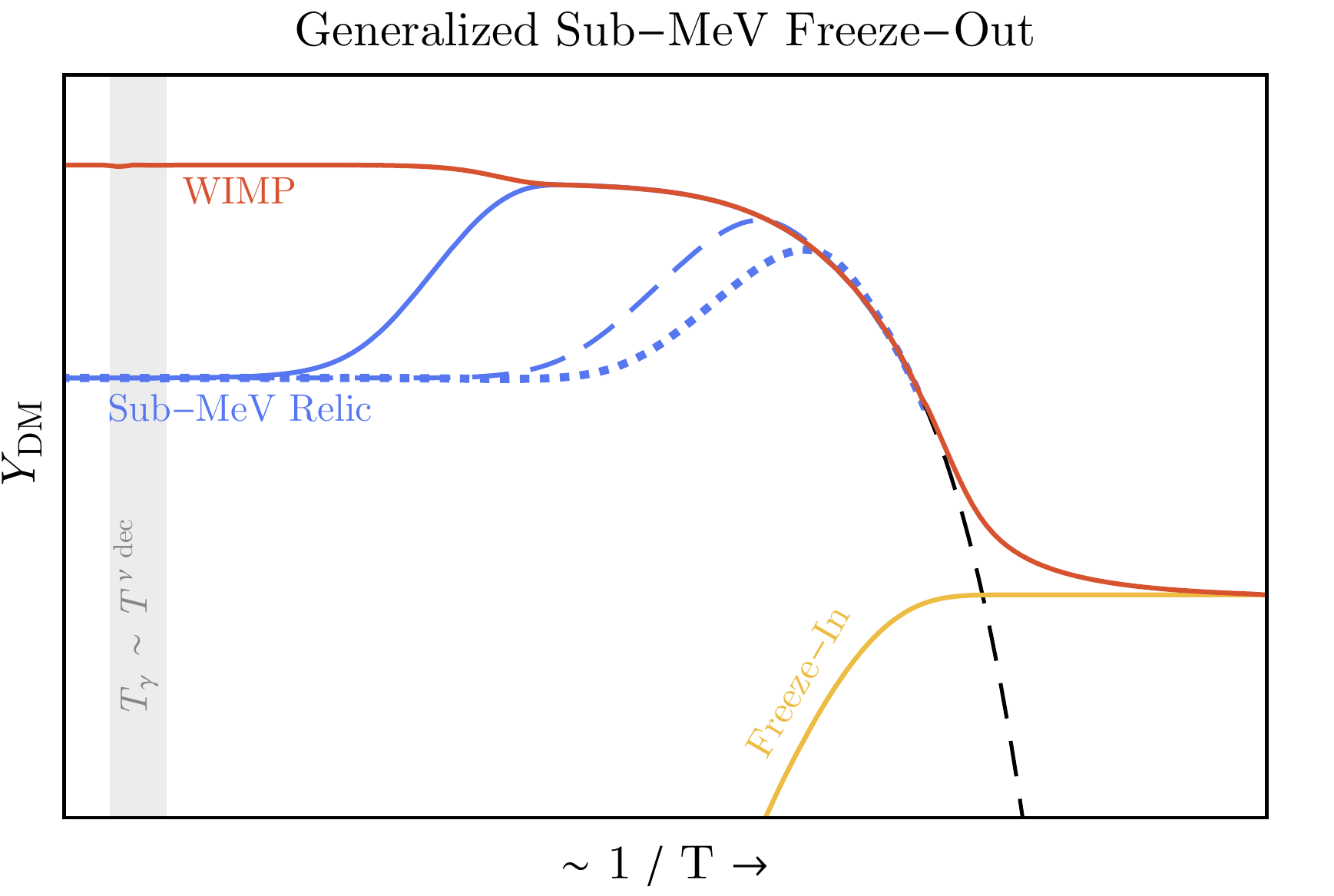} \hspace{-0.5cm}
\caption{Schematic evolution of the dark matter comoving number density ($Y_\text{DM}$) as a function of the photon temperature ($T$).
Compared to Fig.~\ref{fig:freezeout}, we additionally include scenarios in which dark matter equilibrates fully with the Standard Model bath after neutrino-photon decoupling while semi- or non-relativistic (dashed and dotted blue). Such cosmologies interpolate between the two extremes of WIMP-like freeze-out and freeze-in.}
\label{fig:partial}
\end{figure}

A few of these cosmological scenarios are shown in Fig.~\ref{fig:partial}. The blue lines correspond to models in which DM fully equilibrates with the SM neutrino bath after neutrino-photon decoupling but well before thermal freeze-out. The cosmology denoted by the solid blue line was already discussed in detail in Sec.~\ref{sec:temp_and_neff_delequil}, in which the DM is relativistic during HS-SM equilibration. This case is most analogous to the WIMP paradigm, and simple analytic approximations for the evolution of the HS/neutrino temperatures and $\Neff$ were derived in Sec.~\ref{sec:temp_and_neff_delequil}. 
If the HS and neutrino baths equilibrate while DM is semi- or non-relativistic, $\rho_{\nu + X} a^4$ is no longer conserved.
Instead, the system of Boltzmann equations in Eq.~(\ref{eq:Boltzmann1}) must be solved numerically. Such models are shown as the dashed and dotted blue contours in Fig.~\ref{fig:partial}. 

\begin{figure}[t]
\hspace{-0.5cm}
\includegraphics[width=0.6\textwidth]{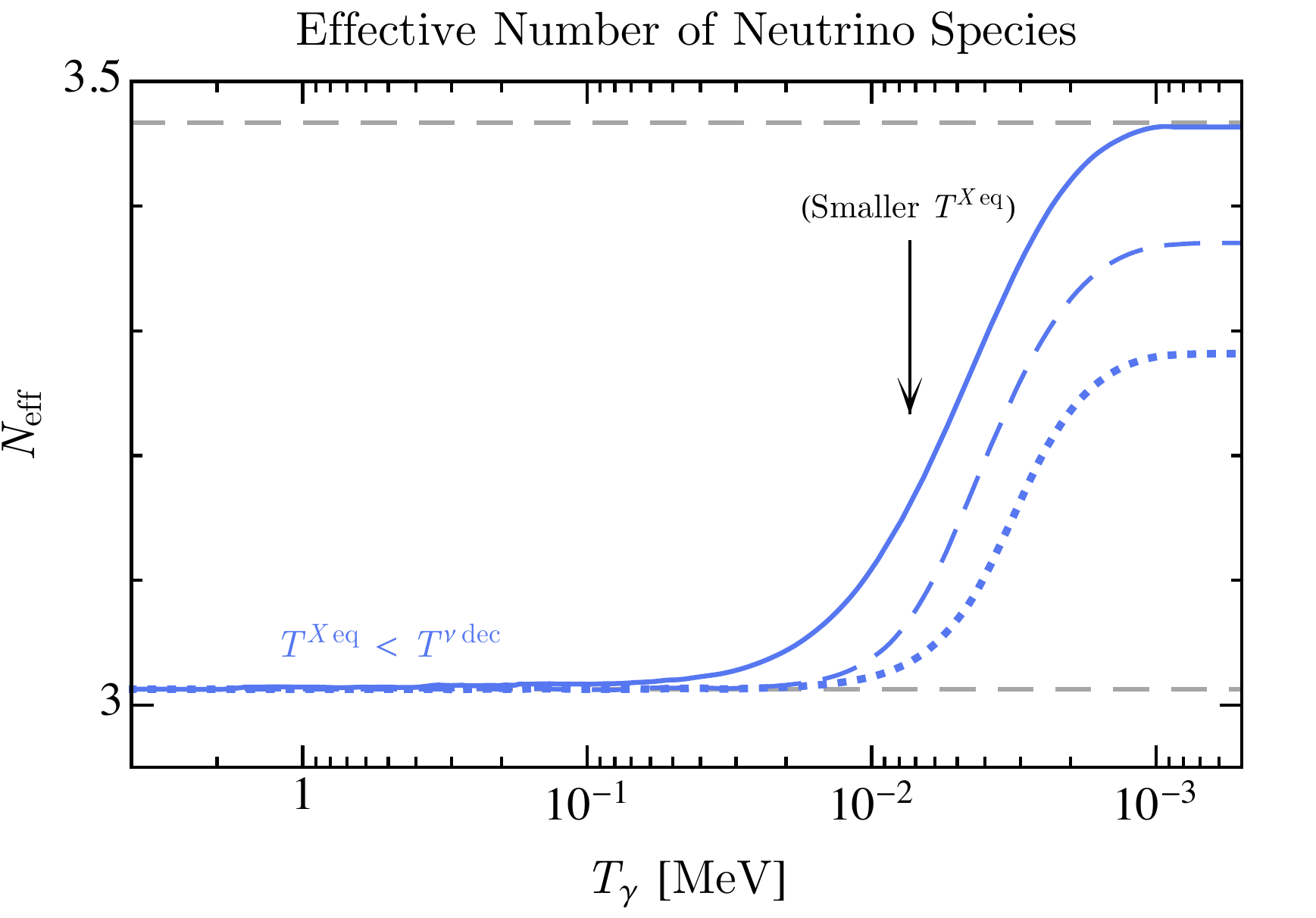} \hspace{-0.5cm}
\caption{The evolution of the effective number of neutrino species in the case that dark matter equilibrates with  neutrinos after neutrino-photon decoupling. The solid (dashed and dotted) contour corresponds to the scenario shown in Fig.~\ref{fig:partial}, where the hidden sector equilibrates with the neutrino bath while the dark matter is relativistic (semi- or non-relativistic). The relativistic case is identical to the one shown in the right panel of Fig.~\ref{fig:equil1}. For concreteness, we have taken the hidden sector to be made up of a 10 keV Majorana fermion and a 5 keV real scalar.}
\label{fig:Neffpartial}
\end{figure}

We show the temperature evolution of $\Neff$ for these generalized scenarios in Fig.~\ref{fig:Neffpartial}, analogous to the right panel of Fig.~\ref{fig:equil1}. The various contours correspond to the examples shown in Fig.~\ref{fig:partial}. For each of these lines in Fig.~\ref{fig:Neffpartial}, HS-SM equilibration occurs after neutrino-photon decoupling. The solid blue contour corresponds to HS-SM equilibration while the DM is relativistic, as studied in Sec.~\ref{sec:temp_and_neff_delequil}. For the dashed and dotted blue contours, equilibration occurs instead when the DM is semi- or non-relativistic, as illustrated in Fig.~\ref{fig:partial}. In Sec.~\ref{sec:temp_and_neff_delequil}, we noted that the increase in $\Neff$ at late times is due to an irreducible heating of the neutrino bath since the equilibration of two initially decoupled gases leads to an overall increase in the comoving entropy of the $\nu-X$ system, i.e.,
\be
dS_{\nu + X} = d Q \, \left( \frac{1}{T_X} -\frac{1}{T_\nu} \right) > 0
\, ,
\ee
where $Q$ is the heat exchanged between the two sectors. If the HS is equilibrated to semi- or non-relativistic temperatures, instead of relativistic ones, the overall heat transfer and entropy increase are reduced, leading to a corresponding decrease in the overall heating of the neutrino bath once the HS becomes non-relativistic. As a result, modifications to $\Neff$ at late times are suppressed compared to relativistic equilibration, as shown explicitly in  Fig.~\ref{fig:Neffpartial}. Although it is beyond the scope of this study, such models constitute an interesting possibility for light, predictive, thermal-like DM. In the next section and the remainder of this work, we will instead focus on an  explicit realization of the cosmological scenarios involving relativistic equilibration, as discussed in Sec.~\ref{sec:temp_and_neff_delequil}.

%

\section{Sub-MeV Dark Matter with a Majoron Mediator}
\label{sec:model1}

The measurement of neutrino oscillations has firmly established the presence of neutrino masses and mixing amongst the different flavor eigenstates. Along with the gravitational observations of DM, the discovery of neutrino masses strongly motivates the existence of physics beyond the SM. We now outline a minimal model that realizes the mechanism described in the previous sections. This model generates the neutrino mass splittings and mixing angles, along with the parameters of the DM sector, through the spontaneous breaking of lepton number. In Sec.~\ref{sec:neutrino}, we discuss the basic framework that is needed to generate the appropriate parameters in the neutrino sector. In Sec.~\ref{sec:dm}, we extend the model to include a stable neutral lepton, which will play the role of DM. We briefly discuss the details of the Higgs sector in Sec.~\ref{sec:scalar}. A more detailed discussion concerning the explicit forms for the masses and interactions of the HS particles is given in Appendix~\ref{sec:majoron_int}.

\subsection{Neutrino Sector}
\label{sec:neutrino}

The SM lacks the necessary ingredients to explain the observed neutrino masses and mixing angles. A simple solution is to include the dimension-five Weinberg operator, $(L H)^2 / \Lambda_\text{uv}$~\cite{Weinberg:1979sa}. Below the scale of electroweak symmetry breaking, this operator generates neutrino masses parametrically of the form $m_\nu \sim v^2 / \Lambda_\text{uv}$, where $v \simeq 246 \GeV$ is the SM Higgs vacuum expectation value (vev) and $\Lambda_\text{uv}$ is the effective scale of new physics. A natural microscopic realization of this operator is the so-called seesaw mechanism, which introduces right-handed neutrinos that are uncharged under the SM gauge group~\cite{Schechter:1980gr,GellMann:1980vs,Mohapatra:1979ia,Yanagida:1979as,Minkowski:1977sc}. 
If neutrinos are Majorana, then $m_\nu \neq 0$ breaks lepton number, $U(1)_L$. The global $U(1)_L$ symmetry can be broken explicitly, as in minimal seesaw models with an explicit Majorana mass for the right-handed neutrinos, or spontaneously when a $U(1)_L$-charged scalar acquires an expectation value.
In the latter case, right-handed neutrino masses are generated dynamically, and the seesaw mechanism can be implemented. Such models involve majorons, the pseudo-Nambu-Goldstone bosons (pNGBs) of $U(1)_L$~\cite{Chikashige:1980ui,Gelmini:1980re,Schechter:1981cv}. This light pseudoscalar will play the role of the mediator between the visible and dark sectors.

In writing down the model, we follow the notation and conventions of Refs.~\cite{Pilaftsis:1993af} and \cite{Garcia-Cely:2017oco}. 
We introduce a complex scalar, $\sigma$, of lepton number $L = 2$,
\be
\sigma = \frac{1}{\sqrt{2}} ~ \left( f + S + i \, J \right)
~,
\ee
where we have assumed that $\sigma$ acquires a non-zero vev, $\langle \sigma \rangle = f / \sqrt{2}$. $S$ and $J$ are the real and imaginary excitations of $\sigma$, where $J$ (often dubbed the majoron) is the Goldstone boson of spontaneous $U(1)_L$-breaking. In the presence of suppressed terms that softly break lepton number, $J$ is a pseudo-Goldstone and acquires a small mass.
Soft $U(1)_L$-breaking terms can arise in the scalar potential, which is examined in Sec.~\ref{sec:scalar} and Appendix~\ref{sec:scalar_masses}.
While we naturally expect $m_J \ll f$, we will not specify the exact form of $U(1)_L$-breaking and treat the majoron mass, $m_J$, as a free parameter of the low-energy theory. A discussion of how such masses may arise from gravitational effects in a more complete theory is provided in Appendix~\ref{sec:planck}.

We introduce three generations of right-handed neutrinos, $N$, with lepton number $L = -1$. The most general renormalizable and $U(1)_L$-symmetric Lagrangian coupling $\sigma$ and $N$ to the SM lepton sector is then given by
\be
\label{eq:Lag1}
- \mathscr{L} \supset y_\nu \, L \, N \, H + \frac{1}{2}\, y_N\, \sigma \, N^2 + \text{h.c.}
~,
\ee
where two-component spinor and flavor indices are implied. Above, $L$ and $H$ are the SM lepton and Higgs doublets, respectively. Below the scale of electroweak and $U(1)_L$-breaking, the interactions in Eq.~(\ref{eq:Lag1}) give rise to the neutrino mass matrix in the $(\nu , N)$ basis,
\be
M_{\nu N} = \begin{pmatrix} 0 & m_D \\ m_D^T & M_N \end{pmatrix}
~,
\ee
where $m_D \equiv y_\nu \, v / \sqrt{2}$ and $M_N \equiv y_N \, f / \sqrt{2}$ are $3 \times 3$ mass matrices.
Diagonalizing $M_{\nu N}$ gives rise to the neutrino mass basis, $n_i$ ($i = 1,2,\dots,6$), with masses $m_i$.
We define the unitary matrix $V$ that diagonalizes the full active-sterile neutrino mass matrix by
\be
V^\dagger \, M_{\nu N} \, V^* = \text{diag}(m_1 , \dots , m_6 ) 
~,
\label{eq:V1}
\ee
where $V$ relates the gauge and mass eigenstates.\footnote{We have chosen to work in the convention where the complex conjugate of $V$ relates the two bases of left-handed Weyl spinors, in accordance with the four-component conventions of Refs.~\cite{Pilaftsis:1993af} and \cite{Garcia-Cely:2017oco}.} 

In the seesaw limit ($m_D / M_N \ll 1$), $n_{1,2,3}$ and $n_{4,5,6}$ are SM-like and sterile-like neutrino species, respectively, with masses schematically of the form $m_{1,2,3} \sim m_D^2 / M_N$ and $m_{4,5,6} \sim M_N \sim f$. The off-diagonal entries of $V$ correspond to 
active-sterile mixing and are suppressed by $m_D / M_N \sim \sqrt{m_{1,2,3}/m_{4,5,6}} \ll 1$. This is made explicit by the Casas-Ibarra parametrization 
as discussed in Appendix~\ref{sec:app1}~\cite{Casas:2001sr}. The interactions of the neutrino mass eigenstates ($n_i$) with the scalar degrees of freedom take the parametric form
\be
- \mathscr{L} \sim (m_i \, m_j)^{1/2} \left( \frac{S + i J}{f} + \frac{h}{v} \right) n_i \, n_j + \text{h.c.}
~,
\ee
where $h$ is the SM Higgs field. The explicit forms of these couplings, along with ones involving SM gauge bosons, are given in Appendix~\ref{sec:app1}. The most important feature of the above interactions is their proportionality to the neutrino masses, which is characteristic of the Higgs mechanism. 
In general, there may be other contributions to the masses of the sterile neutrinos, for instance originating from Dirac masses with additional $L = +1$ sterile neutrinos. In this case, the mass parameters $m_{4,5,6}$ written in these interactions are implicitly assumed to be the piece given by the scale $f$, i.e., $\sim f \times \partial m_{4,5,6} / \partial f$. However, it is important to keep in mind that $M_N \gg f$ is still possible in extended models. We will return to this point later in Sec.~\ref{sec:equil}.

Mass-mixing in the neutrino sector also induces interactions of the sterile states with electroweak currents and generates couplings of $S$ and $J$ to 
charged leptons and quarks  via neutrino loops. These interactions are typically too small to be phenomenologically relevant, but we discuss them briefly in Secs.~\ref{sec:scalar} and \ref{sec:stellar} as well as in Appendix~\ref{sec:majoron_int}.

\subsection{Dark Matter Sector}
\label{sec:dm}
The model described in the previous section involves a viable mechanism for neutrino mass generation. 
  The new particles include a naturally light pseudo-Nambu-Goldstone boson, $J$, that 
  couples to neutrinos. This is precisely the setup required to realize a viable cosmology for 
  sub-MeV DM as described in Secs.~\ref{sec:submev} and \ref{sec:delequil}.
  To complete the model, we introduce an additional Weyl fermion, $\x$, of lepton number $L = -1$ and charged under an additional $\mathbb{Z}_2$. The $\mathbb{Z}_2$ prevents $\x$ from mass-mixing with the active or sterile neutrinos and stabilizes $\x$, which will serve as our DM candidate. The only renormalizable term consistent with the above symmetries is
\be
- \mathscr{L} \supset \frac{1}{2} \,\lambda_\x \,\sigma \, \x^2 +\text{h.c.}
\label{eq:ds_yukawa}
\ee
The phase of $\x$ can be chosen such that the Yukawa coupling, $\lambda_\x$, is purely real. Below the scale of $U(1)_L$-breaking, $\x$ acquires a mass,
\be
\label{eq:DMmass1}
m_\x = \frac{\lambda_\x \, f}{\sqrt{2}}
~.
\ee
In four-component notation, the interactions of the Majorana fermion, $\x$, with $J$ and $S$ are given by
\be
\mathscr{L} \supset ~ \lambda_S^\x ~ S ~ \overline{\x} \, \x + \lambda_J^\x ~ J ~ \overline{\x} \, i \gamma^5 \x
~,
\ee
where the couplings are defined as
\begin{align}
\lambda_S^\x &\equiv - \, \frac{\lambda_\x}{2 \sqrt{2}}  = - \, \frac{m_\x}{2 f}\, ,
\nl
\lambda_J^\x &\equiv \frac{ \lambda_\x }{2 \sqrt{2}} = \frac{m_\x}{2 f}
~.
\end{align}
%

\subsection{Scalar Sector}
\label{sec:scalar}
The $U(1)_L$-preserving renormalizable scalar potential is given by
\be
V_L (H, \sigma) = - \mu_H^2 \, |H|^2 + \lambda_H \, |H|^4 - \mu_\sigma^2 \, |\sigma|^2 + \lambda_\sigma \, |\sigma|^4  + \lambda_{\sigma H} \, |\sigma|^2 |H|^2
~.
\ee
This potential does not generate a mass for the majoron, $J$. However, soft $U(1)_L$-breaking terms such as 
\be
V_{\slashed{L}} = - \left( \mu_\sigma^{\prime \, 2} \, \sigma^2 + a_\sigma \, \sigma \, |H|^2 
+  \text{h.c.} \right)
\ee
can give rise to a radiatively-stable mass for $J$.
The full potential is then given by
\be
V= V_L+ V_{\slashed{L}}
~.
\label{eq:full_pot}
\ee
We fix the phase of $\sigma$ such that its vev, $f$, is real, leaving a single physical phase 
in the couplings $\mu_\sigma^\prime$ and $a_\sigma$. This phase leads to CP-violating mixing of $J$ with $S$ and $h$.
The details of mass-diagonalization and constraints on the scalar potential parameters are discussed in 
Appendix~\ref{sec:scalar_masses}.

As we will illustrate in Sec.~\ref{sec:cosmo}, delayed equilibration of the majoron sector is achieved for $m_J \ll m_S \ll m_h$.
We will assume that the mixing angles in
the scalar sector are small, such that they do not significantly impact physics
in the DM sector. Indeed, we will show in Sec.~\ref{sec:stellar} and in Appendix~\ref{sec:majoron_int} that the Higgs mixing with light states is strongly constrained by 
stellar cooling, rare meson decays, and Higgs decays, implying that the scalar mixing angles are suppressed. In this hierarchical limit, the scalar mass eigenstates ($\p_{1,2,3}$)
are nearly aligned with the gauge basis ($J , S, h$), with masses
\begin{align}
m_1^2 &\simeq m_J^2 \simeq 4 \, \text{Re} \, \mu_\sigma^{\prime \, 2} + \text{Re} \, a_\sigma \, v^2 / \sqrt{2} \, f
\nl
m_2^2 &\simeq m_S^2 \simeq 2 \, \lambda_\sigma \, f^2
\nl
m_3^2 &\simeq m_h^2 \simeq 2 \, \lambda_H \, v^2
~.
\end{align}
This assumption will be relaxed in Sec.~\ref{sec:direct} when we
consider possible signals in futuristic low-threshold direct detection
experiments. 
  For $\lambda_\sigma \sim \mathcal{O}(1)$, the mass of the CP-even scalar, $S$, is near the scale of $U(1)_L$-breaking, $m_S\sim f$. 
  For simplicity, we will fix $m_S = f$ in estimates and numerical results below.

We also note that tree-level mixing between $J$, $S$, and the SM Higgs,
$h$, is not solely responsible for interactions between the HS and the
electrically charged SM fermions. Additional contributions arise from
diagrams involving loops of active/sterile neutrinos and electroweak gauge bosons. We
will not discuss these contributions in detail and instead refer the interested
reader to the relevant sections of Refs.~\cite{Pilaftsis:1993af} and
\cite{Garcia-Cely:2017oco}. For instance, the radiatively induced Yukawa couplings
between $J$, $S$ and the SM quarks and charged leptons are naturally of size,
\be
\label{eq:loopmixing}
\mathscr{L} \sim \frac{m_\nu \, m_f}{16 \pi^2 \, v^2} ~ \left(S ~ \bar{f} f + J ~ \bar{f} i \gamma^5 f \right)
~,
\ee
where $f$ is a charged SM fermion. The effect of this coupling is analogous to $S-h$ and $J-h$ mass-mixing with an effective angle, $\theta_{\rm eff}$, given by
\be
\sin{\theta}_\text{eff} \sim \frac{m_\nu}{16 \pi^2 \, v} \sim \order{10^{-15}}
~,
\ee
where we have taken $m_\nu \sim 0.1 \eV$. As a result, tree-level contributions to $J,S-h$ mixing are only phenomenologically relevant for $\sin{\theta} \gtrsim 10^{-15}$. As we will discuss below, the suppressed size of these radiative interactions makes them irrelevant for the physics governing early universe cosmology and the signals discussed in Secs.~\ref{sec:cosmo} and \ref{sec:signal}. We will come back to these couplings in Sec.~\ref{sec:stellar}, where we discuss effects of $J$ and $S$ on the physics of stellar cooling. 

\section{Cosmology}
\label{sec:cosmo}

\subsection{Equilibration}
\label{sec:equil}

%
\begin{figure}[t]
\hspace{-0.5cm}
\includegraphics[width=0.95\textwidth]{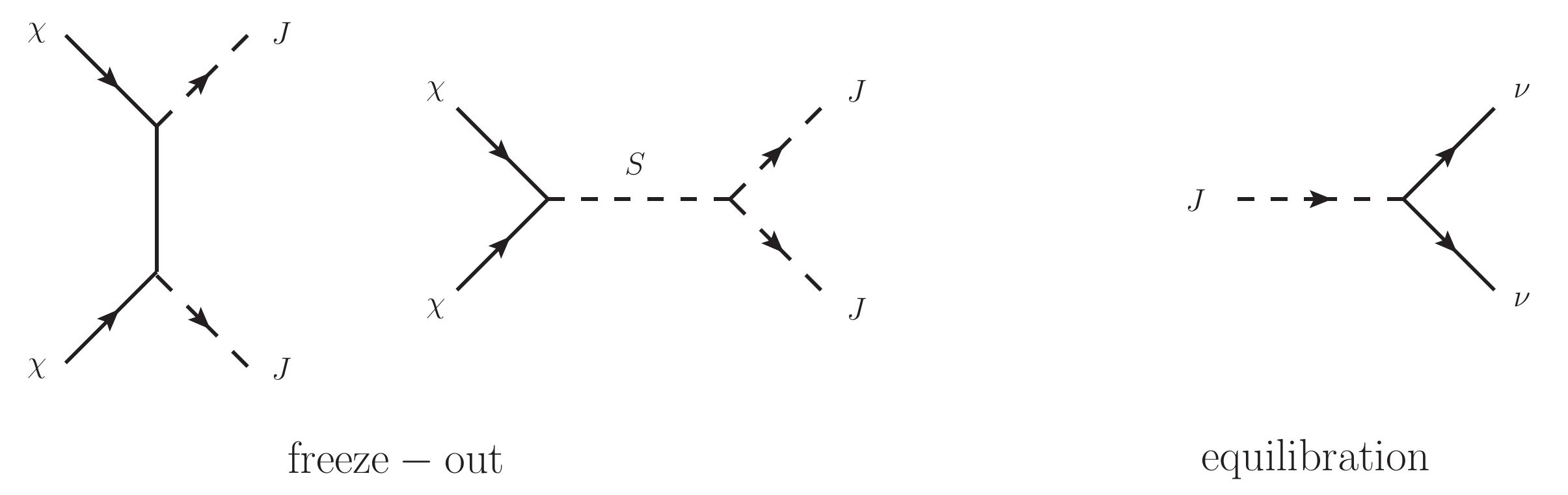} \hspace{0.1cm}
\caption{Representative Feynman diagrams responsible for dark matter freeze-out (left) and equilibration (right). 
}
\label{fig:feynman}
\end{figure}

In this section, we will discuss aspects related to the equilibration of DM with the SM. DM, $\x$, is assumed to equilibrate with the SM neutrinos, $\nu$, while both sectors are relativistic. Since the majoron, $J$, is a pseudo-Goldstone of $U(1)_L$, we naturally take $m_\x \propto f \gtrsim m_J$. In this case, $\x$ freezes out through annihilations into pairs of on-shell majorons, $\x \x \to J J$, followed by $J \to \nu \nu$, as shown in Fig.~\ref{fig:feynman}. From the interactions given in Sec.~\ref{sec:dm}, the non-relativistic cross section for this process is
\begin{align}
\label{eq:xxjj}
\sigma v (\x \x \to J J) &\simeq \vdm^2 ~ \frac{m_\x^2}{64 \pi \, f^4} ~ \frac{(1 - r_J^2)^{1/2}}{(1-r_J^2/2)^4} ~ \left( 1 - 2 \, r_J^2 + \frac{4}{3} \, r_J^4 - \frac{1}{3} \, r_J^6  + \frac{1}{32} \, r_J^8 \right)
~,
\end{align}
where $\vdm$ (not to be confused with the SM Higgs vev) is the relative DM velocity, and we have defined the mass ratio $r_J \equiv m_J / m_\x < 1$. In Eq.~(\ref{eq:xxjj}), we have also taken the limit that $m_S \simeq f \gg m_\x , m_J$. This form suggests that $\x$ acquires an abundance in agreement with the observed DM energy density for
\be
\label{eq:miracle1}
m_\x \sim \order{10^2} ~ \frac{f^2}{(T^\text{MRE} \, \mpl)^{1/2}}
~.
\ee
Hence, $f \sim 10 \MeV - 1 \GeV$ for $m_\x \sim \text{keV} - \text{MeV}$. 
In the minimal model described in Sec.~\ref{sec:model1}, the masses of the sterile neutrinos and HS scalars are also governed by the $U(1)_L$-scale, $f$, and therefore, we parametrically expect $M_N \sim m_S \sim f \lesssim \text{GeV}$.

These parametric estimates for the relevant mass scales suggest that processes involving $N$, $J$, and $S$ are all potentially relevant when considering equilibration between the DM and SM sectors. We will assume that rates for scattering processes in the HS, such as $\x \x \leftrightarrow J J$, are large compared to reactions involving both HS and SM species. Therefore, equilibration between the SM and a single species in the HS rapidly equilibrates all of the lightest particles in the  HS, namely $\x$ and $J$. As noted in Sec.~\ref{sec:delequil}, sub-MeV thermal relics are viable provided that the HS equilibrates with the SM at temperatures below $\Tnu \sim 2 \MeV$. Therefore, it is imperative that processes involving SM neutrinos and $N$, $J$, and $S$ do not equilibrate 
before this point. We now proceed to discuss these various processes in detail.

In the limit that $m_{J,S} \gg \text{eV}$, the decay rates of $J$ and $S$ into SM neutrinos are 
\begin{align}
\label{eq:Jdecay1}
\Gamma (J \to \nu \, \nu) &\simeq \frac{m_J}{16 \pi \, f^2} ~ \sum\limits_{i = 1-3} m_i^2
\nl
\Gamma (S \to \nu \, \nu) &\simeq \frac{m_S}{16 \pi \, f^2} ~ \sum\limits_{i = 1-3} m_i^2
~,
\end{align}
where the sum is over the three active neutrino flavors. 
From examining the Boltzmann equations in Eq.~(\ref{eq:Boltzmann1}), the effective energy transfer rates from decays and inverse-decays that can be compared to Hubble expansion are 
\begin{align}
\label{eq:Jdecay2}
\Gamma^{X \text{ eq}} (J \leftrightarrow \nu \, \nu) &\simeq \frac{m_J \, n_J^\text{eq} (T_\nu)}{\rho_\nu^\text{eq}(T_\nu)} ~  \Gamma (J \to \nu \, \nu)
\nl
\Gamma^{X \text{ eq}} (S \leftrightarrow \nu \, \nu) &\simeq \frac{m_S \, n_S^\text{eq} (T_\nu)}{\rho_\nu^\text{eq} (T_\nu)} ~  \Gamma (S \to \nu \, \nu)
~,
\end{align}
where $n_{J,S}^\text{eq}$ is the equilibrium number density of $J$, $S$, respectively~\cite{Adshead:2016xxj,Csaki:2017spo}. These processes are able to maintain kinetic equilibrium between the HS and SM if $\Gamma^{X \text{ eq}} (J, S \leftrightarrow \nu \, \nu) \gtrsim H$. 
The ratio,
\be
\Gamma^{X \text{ eq}} (J ,S \leftrightarrow \nu \nu) \, / \, H
\, ,
\ee
peaks at temperatures comparable to the mass of the decaying particle, $T \sim m_{J,S}$. For concreteness, let us assume that $m_S \gtrsim m_\x \gtrsim m_J$. We find that equilibration occurs at temperatures $T_\nu \gtrsim m_\x$ through $J \leftrightarrow \nu \nu$ decays  if 
\be
\label{eq:equildec}
\frac{\Gamma^{X \text{ eq}} (J \leftrightarrow \nu \, \nu)}{H} \Bigg|_{T_\nu \sim m_\x} \sim ~~ \order{1} \times \left( \frac{m_\x}{100 \keV} \right)^{-2} \left( \frac{m_\x}{m_J} \right)^{-2} ~ \gtrsim ~~ 1
~,
\ee
or through $S \leftrightarrow \nu \nu$ decays  if 
\be
\label{eq:equildec2}
\frac{\Gamma^{X \text{ eq}} (S \leftrightarrow \nu \, \nu)}{H} \Bigg|_{T_\nu \sim m_S} \sim ~~ \order{10^{-4}} \times \left( \frac{m_\x}{100 \keV} \right)^{-3/2} ~ \gtrsim ~~ 1
~.
\ee
In the above estimates, we have set $m_\nu \sim 0.1 \eV$, $m_S \sim f$, and have fixed $f$ to the thermally-favored value in Eq.~(\ref{eq:miracle1}). Eqs.~(\ref{eq:equildec}) and (\ref{eq:equildec2}) imply that for $m_\x \sim \text{keV} - \text{MeV}$ and $m_J \gtrsim 10^{-2} \, m_\x$, equilibration through $J \leftrightarrow \nu \nu$ dominates over $S \leftrightarrow \nu \nu$. 

Decays of the sterile neutrinos ($N \leftrightarrow J \nu$) are also potentially able to equilibrate the two sectors. For the simplest choices of mixing parameters ($R = \mathbb{1}$ in Appendix~\ref{sec:app1}), each generation of $N$ couples to a single generation of $\nu$. For $M_N \gg m_J$, the corresponding decay rate is
\be
\Gamma (N \to J \, \nu) \simeq \frac{m_\nu \, M_N^2}{16 \pi \, f^2}
~.
\ee
$\Gamma^{X \text{ eq}} (N \leftrightarrow J \, \nu)$ is given by the analogous form of Eq.~(\ref{eq:Jdecay2}). We find that these decays efficiently equilibrate the DM and SM sectors if
\be
\label{eq:equildec3}
\frac{\Gamma^{X \text{ eq}} (N \leftrightarrow J \, \nu)}{H} \Bigg|_{T_\nu \sim M_N} \sim ~~ \order{10^5} \times \left( \frac{m_\x}{100 \keV} \right)^{-1} ~ \gtrsim ~~ 1
~,
\ee
where, once again, we have fixed $f$ to the thermally-favored value in Eq.~(\ref{eq:miracle1}). If these processes equilibrate the DM and SM sectors before neutrino-photon decoupling (which is possible if $M_N \gtrsim \text{few} \times 100 \MeV$), then $N \leftrightarrow J \nu$ decouples above the QCD phase transition, resulting in $\xiDMinitial \sim (10/100)^{1/3} \sim 0.5$. From Fig.~\ref{fig:Neff_Analytic}, such values of $\xiDMinitial$ still significantly alleviate the bounds from measurements of $\Neff$. However, as we will see below in a detailed calculation, thermal freeze-out of $\x$ often favors $M_N \sim f \lesssim \text{few} \times \order{100} \MeV$ and hence potentially larger values of $\xiDMinitial$, worsening this scenario to some degree. To summarize, in the minimal models considered so far, sterile neutrino decays ($N \leftrightarrow J \nu$) often (but not always) prematurely equilibrate the DM and SM sectors, spoiling the mechanism of Sec.~\ref{sec:delequil}.

These issues can be circumvented, for instance, if the mass of $N$ has contributions from additional heavier scales ($M_N \gg f$) or if the post-inflation reheat temperature of the universe is comparatively small ($\text{MeV} \lesssim T_\text{RH} \ll f$). The first case can be realized if the mass of $N$ is lifted by an additional right-handed neutrino, $N^c$, of opposite lepton number, $L= +1$. This charge assignment allows for a Dirac mass involving $N$ and $N^c$ which can be parametrically larger than the scale $f$. 
The second possibility, which involves a low reheat temperature, avoids premature equilibration mediated by on-shell sterile neutrinos with $M_N \sim f$. However, 
processes involving intermediate off-shell sterile neutrinos can still potentially equilibrate the HS and SM bath before neutrino-photon decoupling.
Such reactions include $J \nu \leftrightarrow J \nu$ through an intermediate off-shell $N$. 
In the limit that $m_J \ll T \ll M_N$, the cross section is parametrically of size
\be
\sigma v (J \nu \to J \nu) \sim \frac{m_\nu^2}{f^4}
~.
\ee
After fixing $f$ to the cosmologically-favored value in Eq.~(\ref{eq:miracle1}), this implies that $J \nu \leftrightarrow J \nu$ never maintains equilibrium between DM and the SM for $T_\text{RH} \lesssim \text{TeV} \times (m_\x / 100 \keV)^2$.

Other scattering processes include $\x \nu \leftrightarrow \x \nu$ through $J$ and $S$ exchange, $J \nu \leftrightarrow Z \nu$, and $S t \leftrightarrow h t$, where $t$ is the SM top quark. We find that the rates of equilibration for these reactions are subdominant compared to the ones considered above since they are suppressed by additional small couplings.\footnote{For $m_J \ll \text{eV}$, $\x \nu \leftrightarrow \x \nu$ through $J$ exchange may dominate over $J \leftrightarrow \nu \nu$. However, a simple estimate using Eq.~(\ref{eq:wimp1}) shows that, in this case, equilibration for $m_\x \gtrsim \order{\text{keV}}$ is only possible for neutrino masses that are larger than what is experimentally allowed, i.e., $m_\nu \gg \text{eV}$.} The strength of $S t \leftrightarrow h t$ or $J t \leftrightarrow h t$ explicitly depends on the scalar mixing angles defined in Eq.~(\ref{eq:mixingmatrix}). For $T_\text{RH} \gtrsim m_h$, demanding that these processes do not equilibrate the DM and SM sectors at temperatures above a few MeV requires scalar mixing angles smaller than $\order{10^{-8}}$.  On the other hand, for $T_\text{RH} \sim \text{few} \times \text{MeV}$, similar processes, such as $J e \leftrightarrow \gamma e$, do not equilibrate the two sectors for mixing angles less than $\order{10^{-1}}$. 

\subsection{Dark Matter Freeze-Out}
\label{sec:fo}

In Sec.~\ref{sec:equil}, we demonstrated that various processes can potentially equilibrate the DM and neutrino baths at relativistic temperatures ($T \gg m_\x , m_J$) and after neutrino-photon decoupling ($T \lesssim \text{MeV}$). To acquire a relic abundance that is in agreement with the observed DM energy density, $\x$ must remain in chemical equilibrium until it is non-relativistic, freezing out at temperatures $T \sim m_\x / 10$.  As mentioned in the beginning of Sec.~\ref{sec:equil}, DM freeze-out proceeds through annihilations into pairs of on-shell majorons, i.e., $\x \x \to J J$, followed by $J \to \nu \nu$ (see Fig.~\ref{fig:feynman}). For convenience, we repeat the form for the non-relativistic cross section from Eq.~(\ref{eq:xxjj}),
\begin{align}
\label{eq:xxjj2}
\sigma v (\x \x \to J J) &\simeq \vdm^2 ~ \frac{m_\x^2}{64 \pi \, f^4} ~ \frac{(1 - r_J^2)^{1/2}}{(1-r_J^2/2)^4} ~ \left( 1 - 2 \, r_J^2 + \frac{4}{3} \, r_J^4 - \frac{1}{3} \, r_J^6  + \frac{1}{32} \, r_J^8 \right)
~.
\end{align}
In calculating the relic abundance of $\x$, we follow the semi-analytic approach as detailed in Refs.~\cite{Berlin:2016gtr} and \cite{Chacko:2015noa},
\be
\label{eq:omegaDM}
\Omega_\x \, h^2 \simeq 8.5 \times 10^{-11} ~ \frac{x_f \sqrt{g_*^\text{eff}}}{g_*^\gamma} \left( \frac{3 \, \xi_X \, b \, / \, x_f}{\text{GeV}^{-2}} \right)^{-1}
~,
\ee
where $\xi_X$ is evaluated at freeze-out, $b \equiv \sigma v / \vdm^2$ as in Eq.~(\ref{eq:xxjj2}), and 
\be
\label{eq:gstareff}
g_*^\text{eff} \equiv g_*^\gamma + g_*^\nu \, \xi_\nu^4 + g_*^X \, \xi_X^4
~.
\ee
As before, $X$ collectively denotes the light species in the HS ($\x$ and $J$). $x_f$ is the value of $x \equiv m_\x / T$ at freeze-out, and can be solved numerically through the relation
\be
\label{eq:xf}
x_f \simeq \xi_X \ln{\left( \frac{c (c+2)}{4 \pi^3} ~ \sqrt{\frac{45}{2}} ~ \frac{2}{\sqrt{g_*^\text{eff}}} ~ m_\x \, \mpl ~ \frac{\xi_X^{5/2} \, 6 \, \xi_X \, b \, / \, x_f}{\sqrt{x_f} \left( 1 - 3 \, \xi_X \, / \, 2 \, x_f\right)} \right)}
~,
\ee
where $c \sim \order{1}$ is a constant chosen by matching to numerical solutions of the Boltzmann equation.

\begin{figure}[t]
\hspace{-0.5cm}
\includegraphics[width=0.6\textwidth]{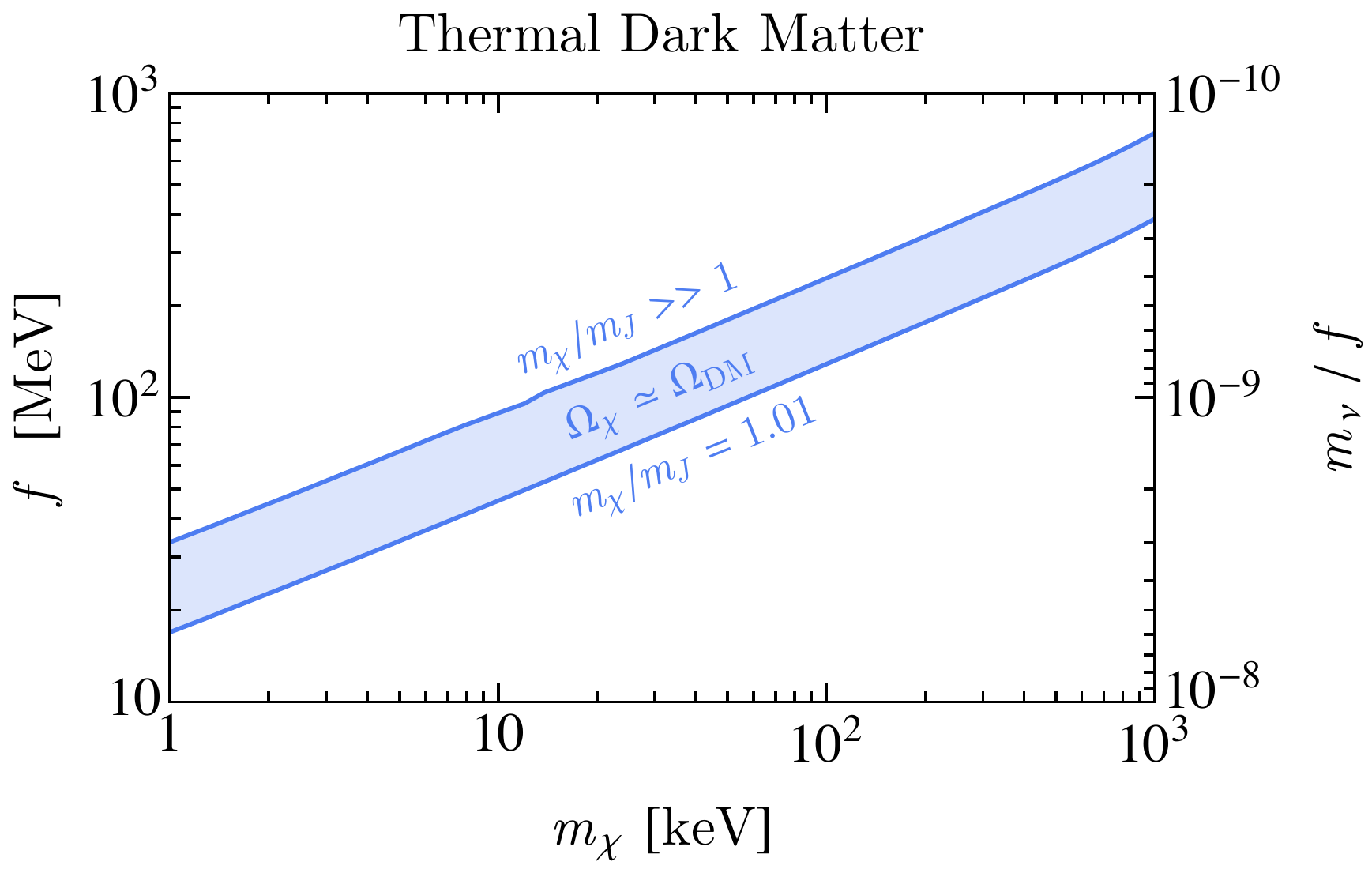} \hspace{0.1cm}
\caption{Values of the scale $f$ required for $\x$ to freeze out with an abundance that is in agreement with the observed dark matter energy density, assuming that the hidden sector is able to equilibrate with the Standard Model while relativistic. The thickness of the band corresponds to varying the $\x-J$ mass ratio between $m_\x / m_J = 1.01$ and $m_\x / m_J \gg 1$. On the right-axis, we also show the ratio $m_\nu / f$, fixing the neutrino masses to $m_\nu = 0.1 \eV$. This ratio is representative of the size of interactions between the Standard Model neutrinos and the majoron, $J$.}
\label{fig:fFO}
\end{figure}

In Fig.~\ref{fig:fFO}, we show the value of $f$ as a function of the DM mass, $m_\x$, that is needed for an adequate freeze-out abundance of $\x$, assuming that the HS equilibrates with the SM neutrinos at relativistic temperatures, i.e., $T \gg m_\x, m_J$. We have taken $m_\x > m_J$, and the thickness of the contour in Fig.~\ref{fig:fFO} corresponds to varying the $\x-J$ mass ratio between $m_\x / m_J = 1.01$ and $m_\x / m_J \gg 1$. In calculating the thermal values of $f$, we have utilized the semi-analytic results in Eqs.~(\ref{eq:omegaDM}) and (\ref{eq:xf}). Note that Fig.~\ref{fig:fFO} is in agreement with the parametric estimate of Eq.~(\ref{eq:miracle1}),
\be
f \sim \order{10}^{-1} ~ m_\x^{1/2} ~ \left( T^\text{MRE} \, \mpl \right)^{1/4}
~.
\ee
These cosmologically-favored values of $f$ imply the presence of new physics associated with the spontaneous breaking of $U(1)_L$ below the GeV-scale. 

\begin{figure}[t]
\hspace{-0.5cm}
\includegraphics[width=0.6\textwidth]{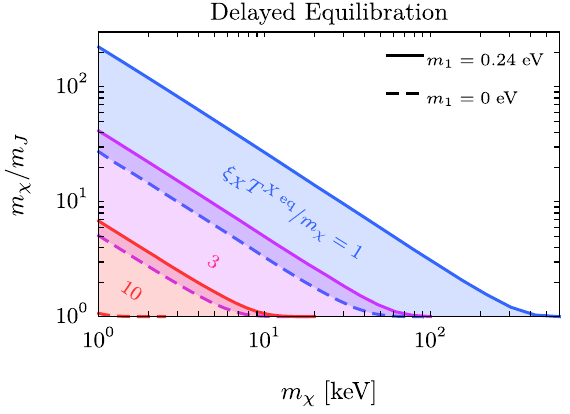} \hspace{-0.5cm}
\caption{The approximate dark matter-majoron mass ratio that is needed for the hidden sector to relativistically (red and purple) or semi-relativistically (blue) equilibrate with the Standard Model neutrino bath. The different colored bands (bounded by solid and dashed lines on top and bottom) correspond to hidden sector temperatures at equilibration of $\xi_X \Tke = (1, 3 , 10 ) \times m_\x$. The width of each band is given by varying the lightest Standard Model neutrino mass, $m_1$, within the cosmologically allowed range of $m_1 = 0 \eV$ (dashed) and $m_1 = 0.24 \eV$ (solid). The scale $f$ is set to the thermal relic value computed in Fig.~\ref{fig:fFO}.
}
\label{fig:fFO2}
\end{figure}

Fig.~\ref{fig:fFO2} shows the required DM-majoron mass ratio, $m_\x / m_J$, as a function of $m_\x$ for various values of the lightest neutrino mass, $m_1$, such that $\x$ acquires an adequate cosmological abundance \emph{and} that the HS relativistically equilibrates with the neutrino bath. In doing so, we fix the scale $f$ as in Fig.~\ref{fig:fFO} and assume that equilibration is dominated by the process $J \leftrightarrow \nu \nu$. In this case, the HS equilibrates relativistically with the SM neutrinos ($\xi_X \Tke \sim \text{few} \times m_\x$) if Eq.~(\ref{eq:equildec}) is fulfilled, which in turn fixes the mass of the majoron, $m_J$, as a function of $f$, $m_1$, and $m_\x$. We are also interested in the generalized scenario of Sec.~\ref{sec:nonrelequil}, in which the HS equilibrates with the SM while $\x$ is semi- or non-relativistic. The different colored regions in Fig.~\ref{fig:fFO2} correspond to HS temperatures at HS-SM equilibration of $ \xi_X \Tke = (1, 3, 10) \times m_\x$. The width of each band is given by varying the lightest neutrino mass within the cosmological allowed range of $m_1 = 0 \eV - 0.24 \eV$~\cite{Ade:2015xua}. After fixing $m_1$, the masses of the other SM neutrinos are given by the observed mass splittings~\cite{Gonzalez-Garcia:2014bfa,Esteban:2016qun}. The regions in Fig.~\ref{fig:fFO2} were obtained by solving the Boltzmann equations in Eq.~(\ref{eq:Boltzmann1}) 
to find $\Tke$. 
The qualitative behavior can also be obtained by comparing the rate of $J \leftrightarrow \nu \nu$ with the Hubble expansion rate.

We conclude this section with a brief derivation of the scaling in Fig.~\ref{fig:fFO2}. 
In Sec.~\ref{sec:equil}, we argued that if the decays and inverse-decays of sterile neutrinos ($N \leftrightarrow J \nu$) are suppressed either through low reheat temperatures or additional contributions to $M_N$, then majoron decays ($J \leftrightarrow \nu \nu$) are dominantly responsible for equilibrating the two sectors below the temperature of neutrino-photon decoupling. Solving Eq.~(\ref{eq:equildec}) for $f$ at $T \sim \Tke$, we find $f^2 \sim m_\nu^2 \, m_J^2 \, \mpl / (\Tke)^3$. Substituting this into Eq.~(\ref{eq:miracle1}) and solving for $m_\x$ gives
\be
\label{eq:miracle2a}
m_\x \sim \left( \frac{m_\x}{m_J} \right)^{-1} ~ \left( \frac{\Tke}{m_\x} \right)^{-3/2} ~ \left( \frac{\mpl}{T^\text{MRE}} \right)^{1/4} ~ m_\nu
\, .
\ee
If we enforce that $\Tke \gtrsim m_\x \gtrsim m_J$, then Eq.~(\ref{eq:miracle2a}) reduces to
\be
\label{eq:miracle2}
m_\x \lesssim
~ \left( \frac{\mpl}{T^\text{MRE}} \right)^{1/4} ~ m_\nu
~ \sim ~ \text{MeV} 
.
\ee
Eq.~(\ref{eq:miracle2}) implies that 
the sub-MeV scale for thermal DM is a natural consequence of the smallness of the observed neutrino masses. This numerical coincidence is surprising, since the MeV-scale has been motivated here in a completely independent manner, compared to the discussion in the beginning of this work. Hence, the framework and model described in the previous sections self-consistently motivate thermal DM below the MeV-scale.

\section{Signals and Constraints}
\label{sec:signal}

\begin{figure}[t]
\hspace{-0.5cm}
\includegraphics[width=0.6\textwidth]{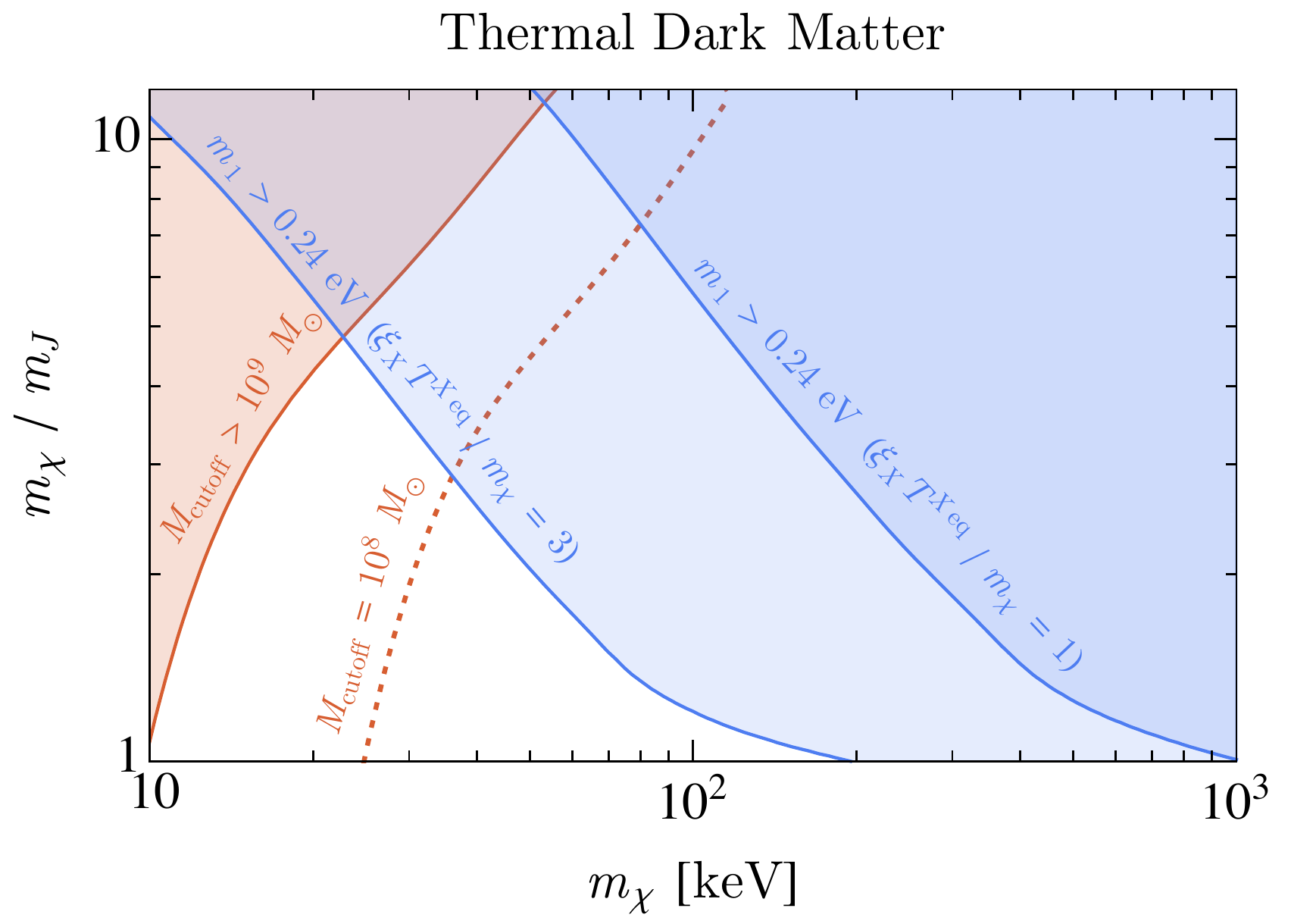} \hspace{-0.5cm}
\caption{
  The viable dark matter parameter space for a sub-MeV hidden sector coupled to Standard Model neutrinos. For every value of the dark matter mass, $m_\x$, and dark matter-majoron mass ratio, $m_\x/m_J$, the lepton number breaking scale, $f$, is fixed to reproduce the correct relic abundance, as in Fig.~\ref{fig:fFO}. Requiring that the hidden sector equilibrates with the neutrino bath at a given temperature
  sets a lower bound on the neutrino masses; in the blue shaded regions, this lower bound exceeds the 
  upper limit on $\sum m_\nu$ set by CMB measurements for $\xi_X \Tke / m_\x = 1,3$. In the red shaded regions, dark matter free-streaming or 
  acoustic oscillations in the hidden sector result in a cutoff in the matter power spectrum that is inconsistent with the smallest 
  observed dark matter substructures. Since the smallest halo mass is subject to uncertainty, we show the resulting 
  constraint for $M_\text{cutoff} = 10^9 \, M_\odot$ (solid red) and $10^8 \, M_\odot$ (dotted red).
  }
\label{fig:model}
\end{figure}

We now discuss signals and constraints for the model outlined in
Secs.~\ref{sec:model1} and \ref{sec:cosmo}. These include cosmological and astrophysical considerations
of the CMB, the small- and large-scale structure of matter, neutrino scattering
in the early universe, DM self-interactions, and stellar cooling. We also
briefly explore the possibility of observing more direct signals in terrestrial
searches for light DM, sterile neutrinos, or majorons. While many of the models are
already tightly constrained by existing measurements, there remain viable
regions of parameter space that will be decisively tested in the near future.
This is illustrated explicitly in Fig.~\ref{fig:model} as a function of the mass ratio, $m_\x / m_J$, and the DM mass, $m_\x$. Throughout
this parameter space, we fix $\xi_X \Tke = (1-3) \times m_\x$, so that the DM sector
equilibrates with the SM before $\x$ is non-relativistic (well before freeze-out), analogous to the standard
picture for thermal WIMPs. We also fix the lightest SM neutrino mass, $m_1$, and the scale
of $U(1)_L$-breaking, $f$, as in Figs.~\ref{fig:fFO} and \ref{fig:fFO2} so that
$\x$ makes up the entire DM abundance at late times. 

%

\subsection{CMB}
\label{sec:cmb}

The general framework discussed in Sec.~\ref{sec:delequil} will be
decisively tested by observations of the CMB in various ways. 
First, the light HS degrees of freedom alter the radiation 
energy density at the time of recombination; this modification is encoded in the effective number of
neutrinos, $\Neff$.
The impact of $\Neff$ on the CMB sky is described in Sec.~\ref{sec:CMB}.
Near-future CMB-S3
and S4 experiments, consisting of a collection of ground-based telescopes, will
have unprecedented sensitivity to deviations of $\Delta \Neff \simeq 0.06$ and $0.027$ within
$1 \sigma$, respectively~\cite{Abazajian:2016yjj}. As noted in
Eq.~(\ref{eq:NeffMin}), the presence of even a single sub-MeV degree of
freedom in the HS that relativistically equilibrates with the SM neutrinos
below an MeV implies that $\Delta \Neff \gtrsim 0.18$ at the time of recombination.
Hence, CMB-S4 experiments will definitively test the presence of such thermal
relics, regardless of their contribution to the abundance of cosmological DM.

The CMB also constrains these models through indirect measurements of SM neutrino masses.
Because the majoron is the pseudo-Goldstone of 
lepton number, its interactions with neutrinos are set by $m_\nu/f$, 
which in turn determines the equilibration temperature, $\Tke$, as described in Sec.~\ref{sec:cosmo} (see Fig.~\ref{fig:fFO2}).
For fixed $m_\x$ and $m_J$, larger HS-SM equilibration temperatures  require heavier SM neutrinos.
Thus, for certain choices of parameters, relativistic equilibration of the HS
can be in conflict with upper bounds on neutrino masses.
One such upper bound comes from Planck measurements of the temperature
power spectrum (TT), which currently constrains the sum of the SM neutrino
masses such that $\sum\limits_{i = 1-3} m_i \lesssim 0.72 \eV$~\cite{Ade:2015xua}.
This corresponds to a bound on the lightest neutrino mass of $m_1 \lesssim 0.24 \eV$ 
for the normal and inverted mass orderings. Combinations of the Planck dataset with other cosmological 
observations further tighten this bound as much as $\sum\limits_{i=1-3} m_i \lesssim 0.18 \eV$~\cite{Ade:2015xua,Capozzi:2017ipn}. However, it has been noted 
that uncertainties in the CMB lensing amplitude can significantly weaken these cosmological 
limits~\cite{Capozzi:2017ipn}. Hence, for simplicity, we show only the Planck TT constraint
in Fig.~\ref{fig:model}, for various choices of the equilibration temperature.

\subsection{Structure Formation}
\label{sec:wdm}

\subsubsection{Dark Matter Free-Streaming and Acoustic Oscillations}
The models considered throughout this work can lead to observable deviations in
the observed matter power spectrum. 
Light DM that remains coupled to HS or SM radiation until late times can suppress power at small scales via two distinct mechanisms: free-streaming and acoustic oscillations. These processes wash out structure below a characteristic comoving length
scale, $\lambda_\text{cutoff}$, which sets a lower bound on the
present day mass of the smallest gravitationally collapsed DM structures, 
\be
\label{eq:mcutoff}
M_\text{cutoff} = \frac{4 \pi}{3} \, \rho_{_\text{DM}} \, \lambda_\text{cutoff}^3 \simeq 1.4 \times 10^8 M_\odot \times \left( \frac{\lambda_\text{cutoff}}{0.1 \text{ Mpc}} \right)^3
~,
\ee
where $\rho_{_\text{DM}} = 1.26 \times 10^{-6} \GeV \text{ cm}^{-3}$
is the present cosmological DM energy density~\cite{Olive:2016xmw}. 
The cutoff scale is determined by solving Boltzmann equations describing the coupled 
DM-radiation system during the epoch of DM decoupling and free-streaming, which modifies the initial 
primordial matter power spectrum~\cite{Hofmann:2001bi,Green:2005fa,Loeb:2005pm}. Here, we merely estimate 
the cutoff scales for the two effects following Refs.~\cite{Boehm:2004th,Gondolo:2012vh,Bertoni:2014mva}.
The scale that enters Eq.~(\ref{eq:mcutoff}) is then given by the larger of the two
lengths associated with free-streaming ($ \lambda_\text{FS}$) and acoustic
oscillations ($ \lambda_\text{AO}$),
\be
\label{eq:cutofflength}
\lambda_\text{cutoff} \simeq \text{max} \left( \lambda_\text{FS},\,\lambda_\text{AO} \right)
\, .
\ee
We now discuss each of these in turn.

Once $\x$ kinetically decouples from the radiation bath (either from HS majorons or SM neutrinos), it begins to freely diffuse across the universe, suppressing matter perturbations smaller than the free-streaming scale, $\lambda_\text{FS}$. This length scale is defined as the comoving distance traversed by DM from the time of decoupling (assumed to occur during radiation domination) until matter-radiation equality,
\be
\label{eq:lambdafs}
\lambda_\text{FS} = c_\text{FS} \, \int_{t_\text{KD}}^{t_\text{MRE}} dt \,  \frac{v_\x}{a}
~,
\ee
where $a$ is the scale factor, $v_\x = p_\x / E_\x$ is the physical velocity of $\x$, $t_\text{KD}$ and $t_\text{MRE}$ are the cosmological times associated with DM kinetic decoupling and matter-radiation equality, respectively, and $c_\text{FS}$ is an $\order{1}$ number. 
There is some ambiguity in $c_\text{FS}$ due to different conventions and  $\mathcal{O}(1)$ factors that appear in the Boltzmann equation treatment of free-streaming~\cite{Loeb:2005pm}. For example, in Ref.~\cite{Gondolo:2012vh}, $c_\text{FS} = 1/2$, while Ref.~\cite{Loeb:2005pm} finds $c_\text{FS} = \pi/(2 \sqrt{6}) \simeq 0.64$. In evaluating $\lambda_\text{FS}$, we take $c_\text{FS} = 1/2$.
To simplify the evaluation of Eq.~(\ref{eq:lambdafs}), let us assume that $\x$ kinetically decouples while non-relativistic at a photon temperature of $T^\text{KD} \ll \order{\text{MeV}}$. In this case, Eq.~(\ref{eq:lambdafs}) can be simplified to
\begin{align}
\label{eq:lambdafs2}
\lambda_\text{FS} &\simeq c_\text{FS} \left( \frac{4 \pi^3}{135} ~ g_*^\text{eff} ~ \frac{T^\text{KD} \, m_\x}{\xi_X} \right)^{-1/2} \, \frac{\mpl}{T_0} ~ \log{\frac{T^\text{KD}}{T^\text{MRE}}}
\nl
&\simeq 0.13 \text{ Mpc} \times c_\text{FS} ~ \xi_X^{1/2} \left( \frac{T^\text{KD}}{\text{keV}} \right)^{-1/2} \left( \frac{m_\x}{100 \keV} \right)^{-1/2}\left( 1 + 0.14 \, \log{\frac{T^\text{KD}}{\text{keV}}} \right)
~,
\end{align}
where $g_*^\text{eff}$ is defined as in Eq.~(\ref{eq:gstareff}), $T_0 \simeq 2.3 \times 10^{-4}$ eV is the present day photon temperature, $T^\text{KD}$ is the temperature of the photon bath at DM kinetic decoupling, and $\xi_X$ and $g_*^\text{eff}$ are evaluated at $T^\text{KD}$.

Density fluctuations of the DM fluid that enter the horizon while DM is kinetically coupled to SM neutrinos and/or relativistic majorons oscillate with the radiation bath, similar to the baryonic acoustic oscillations in the baryon-photon plasma. The amplitude of these modes 
is damped due to their coupling to radiation. As a result, they do not undergo
the usual logarithmic growth during radiation domination~\cite{Loeb:2005pm}. This results in suppressed power on scales smaller than the comoving horizon at decoupling,
\be
\label{eq:lambdaao}
\lambda_\text{AO} = \int_{0}^{t_\text{KD}} dt \, \frac{1}{a} = \frac{1}{a_\kd H_\kd}
~,
\ee
where $a_\text{KD}$ and $H_\text{KD}$ are the scale factor and Hubble parameter at DM kinetic decoupling. Once again assuming that DM kinetic decoupling occurs at temperatures $T^\text{KD} \ll m_\x , \order{\text{MeV}}$, Eq.~(\ref{eq:lambdaao}) is approximately
\begin{align}
\label{eq:lambdaao2}
\lambda_\text{AO} &\simeq \left( \frac{4 \pi^3}{45} \, g_*^\text{eff} \right)^{-1/2} ~ \frac{\mpl}{T^\text{KD} \, T_0}
\nl
&\simeq 0.1 \text{ Mpc} \times \left( \frac{T^\text{KD}}{\text{keV}} \right)^{-1}
~,
\end{align}
where $g_*^\text{eff}$ is evaluated at $T^\text{KD}$, as in Eq.~(\ref{eq:lambdafs2}).

In order to evaluate Eqs.~(\ref{eq:lambdafs2}) and (\ref{eq:lambdaao2}), we
need to determine the photon temperature at kinetic decoupling, $T^\text{KD}$.
The DM, $\x$, chemically decouples when $\x \x \leftrightarrow J J$ freezes out (see Sec.~\ref{sec:fo}), 
but remains in kinetic equilibrium with the SM bath 
directly through $\x \nu\leftrightarrow \x \nu$ or indirectly through $\x J \leftrightarrow \x J$ ($+ ~
J \leftrightarrow \nu \nu$).
Since $\x J \leftrightarrow \x J$ is governed by
the same couplings as $\x \x \leftrightarrow JJ$, the fact that $\x \x
\leftrightarrow JJ$ freezes out at $T_X \sim m_\x / 10$ implies that $\x J
\leftrightarrow \x J$ decouples at $T_X \sim m_J / 10$. For $T_X \sim m_J / 10$
and $m_J \sim m_\x$, the rate for $\x J \leftrightarrow \x J$ is enhanced over
that of $\x \nu \leftrightarrow \x \nu$ by approximately
\be
\label{eq:scattratio}
\frac{n_J^\text{eq} \, \langle \sigma v (\x J \to \x J)\rangle}{n_\nu^\text{eq} \, \langle \sigma v (\x \nu \to \x \nu)\rangle} \sim \order{10^8} \times \left( \frac{m_\x}{100 \keV} \right)^2 \left( \frac{m_\nu}{\text{eV}} \right)^{-2}
~.
\ee
Hence, $\x \nu \leftrightarrow \x \nu$ decouples well before $\x J \leftrightarrow \x J$, and we expect $\x J \leftrightarrow \x J$ to dictate $T^\text{KD}$. In the limit that $m_\x \gg T_X, m_J$, the differential rate for this scattering process is approximately
\be
\label{eq:xjxj}
\frac{d \sigma}{d t} (\x J \to \x J) \simeq \frac{m_\x^2}{4 \pi \, f^4 \, p_J^2}
~,
\ee
where $p_J$ is the momentum of $J$ in the center of mass frame and $t$ is the usual Mandelstam variable.

We follow Refs.~\cite{Bertoni:2014mva,Gondolo:2012vh} in calculating the temperature at kinetic decoupling, $T^\text{KD}$. 
We estimate $T^\kd$ by equating the momentum relaxation rate for $\x J \leftrightarrow \x J$ (denoted by $\gamma$) to the Hubble expansion rate,
\be
\label{eq:kindec1}
\gamma (\x J \leftrightarrow \x J) (T^\text{KD}) = H (T^\text{KD})
~,
\ee
where 
$\gamma (\x J \leftrightarrow \x J)$ is defined as
\be
\gamma (\x J \leftrightarrow \x J) \equiv \frac{1}{6 m_\x \, T_X} \int_0^\infty \frac{d^3 p_J}{(2 \pi)^3} \, f_J \, (1 + f_J) \, \frac{p_J}{\sqrt{p_J^2 + m_J^2}} \int_{-4 p_J^2}^0 dt (-t) ~ \frac{d \sigma}{d t} 
~.
\ee
Above, $f_J$ is the phase-space density of $J$, and $d \sigma / d t$ is as given in Eq.~(\ref{eq:xjxj}). In the non-relativistic limit and taking $f \gg m_\x, m_J$, this becomes
\be
\label{eq:gamma1}
\gamma (\x J \leftrightarrow \x J ) \simeq \frac{4 \, \xi_X^2}{3 \pi^3} ~ \frac{m_\x \, m_J^2 \, T^2}{f^4} \, e^{- m_J / \xi_X T}
~.
\ee
Eqs.~(\ref{eq:kindec1}) and (\ref{eq:gamma1}) allow us to estimate the kinetic decoupling temperature, $T^\text{KD}$, through the relation
\begin{align}
\label{eq:kindec2}
\frac{m_J}{\xi_X \, T^\text{KD}}
&\simeq \ln{\bigg[ \left(\frac{20}{\pi^9 \, g_*^\text{eff}} \right)^{1/2} \, \xi_X^2 ~ \frac{m_\x \, m_J^2 \, \mpl}{f^4}  \bigg]}
\nl
&\simeq 17 + \ln{\bigg[ \left(\frac{m_\x}{100 \keV}\right) \left(\frac{m_\x}{m_J}\right)^{-2} ~ \xi_X^2 \bigg]}
~,
\end{align}
where $g_*^\text{eff}$ and $\xi_X$ are evaluated at $T^\text{KD}$, and in the second equality we have fixed $f$ to the thermally-favored value, as shown in Eq.~(\ref{eq:miracle1}) and Fig.~\ref{fig:fFO}.

The minimum halo mass, $M_\text{cutoff}$, can be calculated using Eqs.~(\ref{eq:mcutoff}), (\ref{eq:cutofflength}), (\ref{eq:lambdafs2}), (\ref{eq:lambdaao2}), and (\ref{eq:kindec2}). Various astrophysical observations, such as Milky Way satellite counts and the Lyman-$\alpha$ absorption lines of distant quasars, constrain $M_\text{cutoff} \lesssim (10^7 - 10^9) ~ M_\odot$, corresponding to $\lambda_\text{cutoff} \lesssim (0.05 - 0.2) ~ \text{Mpc}$ (see, e.g., Refs.~\cite{Strigari:2008ib,Polisensky:2010rw,Vegetti:2012mc,Viel:2013apy,Vegetti:2014lqa,Baur:2015jsy,Garzilli:2015iwa,Irsic:2017ixq,Kim:2017iwr} and references within). We will conservatively demand that $M_\text{cutoff} \lesssim 10^{9} ~ M_\odot$ as shown by the solid red line in Fig.~\ref{fig:model}, although we additionally highlight regions of parameter space in which $M_\text{cutoff} = 10^{8} ~ M_\odot$ as a dotted red line. 

  The minimum halo mass constraint sets a lower limit on the DM mass of  $m_\x \gtrsim (10-50) \keV$, for the thermal relic parameter space shown in Fig.~\ref{fig:model}. This is a stronger bound compared to the often-quoted limit on warm DM~\cite{Viel:2013apy}, which is usually assumed to have decoupled from the SM while relativistic at large temperatures. In the present model, the momentum of $\x$ redshifts less between chemical decoupling 
and matter-radiation equality because $\x$ remains coupled to the radiation bath of $J$ and $\nu$ until late times. As seen in Fig.~\ref{fig:model}, the bound becomes more severe for larger values of $m_\x/m_J$ since $\x$ decouples later (see Eq.~(\ref{eq:kindec2})) as $m_J\rightarrow 0$. Furthermore, for $m_\x / m_J \lesssim \text{few}$, the cutoff in the power spectrum ($\lambda_\text{cutoff}$) is controlled by free-streaming, while for larger values of $m_\x / m_J$, acoustic oscillations in the HS dominate. This can be understood by taking the ratio of Eqs.~(\ref{eq:lambdafs2}) and (\ref{eq:lambdaao2}). For $m_\x \sim \order{10} \keV$, we find
\be
\frac{\lambda_\text{FS}}{\lambda_\text{AO}} \sim \order{10} \times \left( \frac{m_\x }{T^\text{KD}} \right)^{-1/2} \sim \text{few} \times \left( \frac{m_\x}{m_J} \right)^{-1/2}
~,
\ee
where in the second equality we have used Eq.~(\ref{eq:kindec2}). As a result, acoustic oscillations dominate over free-streaming in controlling the matter power spectrum cutoff for $m_\x / m_J \gtrsim \text{few}$. These limits will be improved in the near-future with, e.g., observations of the 21-cm hydrogen line in the cosmic dark ages~\cite{Sitwell:2013fpa,Sekiguchi:2014wfa,Shimabukuro:2014ava}. For instance, an order of magnitude improvement in the sensitivity to $\lambda_\text{cutoff}$ would probe most of the remaining parameter space in Fig.~\ref{fig:model}.

Various studies have examined the effect of DM-neutrino scattering ($\x \nu \to \x \nu$) on the matter power spectrum~\cite{Boehm:2000gq,Boehm:2004th,Boehm:2014vja,Schewtschenko:2015rno,Boehm:2003xr,Wilkinson:2014ksa,Mangano:2006mp,Serra:2009uu}.
We previously showed in Eq.~(\ref{eq:scattratio}) that this process decouples well before $\x J \leftrightarrow \x J$ 
and therefore is not relevant for structure formation.
However, for completeness we will compare the upper limits derived
in the works listed above to the scattering rate for $\x \nu \leftrightarrow \x
\nu$ in our model. Majoron exchange dominates this process, since $m_J \ll
m_S$; the low-energy cross section takes the parametric form
\be
\label{eq:DMneutrino}
\langle \sigma v \left( \x \nu \to \x \nu \right) \rangle \sim \text{few} \times \frac{m_\nu^2 \, T^4}{f^4 \, m_J^4}
~,
\ee
where the $T^4$ temperature dependence arises from the CP-odd nature of the interaction between the majoron and the non-relativistic $\x$.
For sufficiently large scattering rates, DM and neutrinos are tightly coupled in the early universe, altering the observed matter power spectrum, for instance, in large galaxy surveys. These effects constrain the size of the DM-neutrino opacity, $Q \equiv \langle \sigma v (\x \nu \to \x \nu) \rangle / m_\x$, where the temperature scaling of $Q$ is parametrized as either constant, $Q \propto T^0$, or falling as the temperature squared, $Q \propto T^2$. In the case of constant scaling, the strongest bounds lead to the constraint $Q \lesssim 10^{-33} \text{ cm}^2 / \text{GeV}$~\cite{Wilkinson:2014ksa}. Since the predicted rate in Eq.~(\ref{eq:DMneutrino}) falls as $T^4$, we conservatively compare the upper bound from Ref.~\cite{Wilkinson:2014ksa} to the value predicted in our model at temperatures near matter-radiation equality, $T \sim \text{eV}$, which gives the strongest possible constraint. We find that the predicted rate in our model is many orders of magnitude below this observational limit throughout the relevant parameter space shown in Fig.~\ref{fig:model}. 

\subsubsection{Dark Matter and Neutrino Self-Interactions}

Non-standard neutrino interactions mediated by new forces (such as the
majoron) can also alter the behavior of fluctuations in the photon and baryon
fluids during the early universe.
In the standard cosmology, neutrinos diffuse freely after decoupling 
from the photon plasma at temperatures of a few MeV until they become non-relativistic 
well after recombination.
Such free-streaming radiation creates anisotropic shear stress, which, through
gravity, suppresses the amplitude and shifts the phase of acoustic modes in the
CMB that enter the horizon during this
epoch~\cite{Bashinsky:2003tk,Friedland:2007vv,Hou:2011ec}. However, if
self-interactions (or interactions with another species) allow neutrinos to form a tightly coupled fluid before
matter-radiation equality, the point at which they begin free-streaming is
delayed. As a result, the strength of anisotropic stress is reduced compared to
the SM expectation, and the power in subhorizon fluctuations is correspondingly
increased and shifted in phase towards smaller angular scales. 

Recent studies have investigated the effects of neutrino self-interactions ($\nu \nu \to \nu \nu$) on the CMB, where the strength of the neutrino opacity is parametrized in terms of the dimensionful coefficient of a four-fermion operator, $G_\text{eff}$~\cite{Cyr-Racine:2013jua,Lancaster:2017ksf,Oldengott:2017fhy}. These analyses have found that $G_\text{eff} \lesssim 1 / (50 \MeV)^2$ is consistent with data from Planck, the Sloan Digital Sky Survey, and local measurements of the Hubble parameter. In particular, for the models considered in Sec.~\ref{sec:model1}, elastic neutrino scattering proceeds through the exchange of the light spin-0 mediators, $J$ and $S$. In the limit that $m_\nu \ll \text{eV} \ll m_{J,S}$, the relevant cross section is parametrically
\be
\sigma v (\nu \nu \to \nu \nu) \sim G_\text{eff}^2 \, T_\nu^5
~,
\ee
where the effective coupling is given by
\be
G_\text{eff} \sim \frac{m_\nu^2}{f^2 \, m_{J,S}^2}
~.
\ee
Since $m_J \ll m_S$, elastic neutrino scattering is dominantly governed by majoron exchange, so that $G_\text{eff} \sim m_\nu^2 / ( f^2 \, m_J^2 )$. From Figs.~\ref{fig:fFO}-\ref{fig:model}, the viable parameter space of our model is given by $m_\nu \lesssim 0.1 \eV$, $f \gtrsim 10 \MeV$, and $m_J \gtrsim 100 \eV$, which implies that
\be
G_\text{eff} \lesssim \frac{1}{\left( 10^4 \MeV \right)^2}
~.
\ee
This is orders of magnitude below the upper bound derived in from Refs.~\cite{Cyr-Racine:2013jua,Lancaster:2017ksf,Oldengott:2017fhy,Randall:2007ph}.
We note that the $\nu-J$ coupling in the early universe also delays neutrino free-streaming until $J$ becomes non-relativistic. 
  The bound on delayed free-streaming in Refs.~\cite{Cyr-Racine:2013jua,Lancaster:2017ksf} can be stated in terms of a lower limit on the redshift 
  at neutrino decoupling: $z_{\nu \text{ dec}} > 1.3 \times 10^5$. For the masses $m_J \gtrsim \text{keV}$, as considered in this work, $\nu$ decouples from 
  $J$ well before this epoch.

$J$ and $S$ exchange also gives rise to DM self-scattering ($\x \x \to \x \x$). The self-scattering cross section per DM mass is bounded from observations of the dynamics and structures of galaxy clusters to be $\sigma / m_\x \lesssim \text{cm}^2 / \text{g}$, where the characteristic value of the relative DM velocity is $\vdm^2 \sim 10^{-5}$~\cite{Clowe:2003tk,Markevitch:2003at,Kaplinghat:2015aga}. We follow the discussion in Refs.~\cite{Knapen:2017xzo,Tulin:2012wi,Tulin:2013teo} to calculate the viscosity cross section for the self-scattering of identical DM particles. For $m_S \sim f \gtrsim m_\x$ and in the limit that $v \ll m_J / m_\x \ll 1$, DM self-scattering is dominated by majoron exchange,
\be
\frac{\sigma (\x \x \to \x \x)}{m_\x} \simeq \frac{m_\x}{192 \pi \, f^4}
~.
\ee
For $m_\x \gtrsim \text{keV}$, this rate is maximized for $m_\x \sim \text{keV}$ and $f \simeq 30 \MeV$, where $f$ has been fixed to the thermally-favored value in Fig.~\ref{fig:fFO}. This gives $\sigma (\x \x \to \x \x) / m_\x \lesssim 10^{-6} \text{ cm}^2 / \text{g}$, which is orders of magnitude below the inferred upper bound. 

\subsection{Stellar Cooling}
\label{sec:stellar}

New particles coupled to the SM can lead to additional energy loss mechanisms in stellar systems,
such as supernovae, red giants, and horizontal branch stars. 
One of the most powerful constraints on new light degrees of freedom comes from 
the observed cooling rate of SN1987A~\cite{Raffelt:1996wa}. For $m_J
\lesssim 10 \MeV$, annihilations of SM neutrinos into a light majoron ($\nu \nu
\to J$) can lead to qualitative changes in the measured neutrino burst duration. 
Supernova bounds on majorons have been studied in detail in Refs.~\cite{Choi:1989hi,Kachelriess:2000qc,Farzan:2002wx,Heurtier:2016otg}.
Here we estimate an upper bound on the $J-\nu$ coupling as follows. 
The energy loss rate per unit volume scales as $Q_J \sim m_J \Gamma_J n_\nu$~\cite{Heurtier:2016otg}, where 
$\Gamma_J \sim m_\nu^2 m_J/f^2$ is the zero-temperature majoron decay rate (see Eq.~(\ref{eq:Jdecay1})) and 
$n_\nu$ is the neutrino number density for a given $\nu$ flavor. It is important 
to distinguish between electron and the heavy flavor neutrinos in the core.
The former have a large chemical potential, $\mu_{\nu_e} \simeq 200 \MeV$,
with $n_{\nu_e} \sim \mu_{\nu_e}^3$, while the latter have a thermal population, 
such that $n_{\nu_{\mu,\tau}} \sim T_{\rm SN}^3$, where $T_\text{SN} \sim 30 \MeV$ 
is the core temperature. The larger electron neutrino density 
leads to a stronger constraint on model parameters (unless the 
electron-neutrino-like mass eigenstate is massless).
A conservative bound on the anomalous cooling rate 
is obtained by requiring that the instantaneous majoron-luminosity, $\mathcal{L}_J$, 
does not exceed the total neutrino-luminosity of $\mathcal{L}_\nu = 3\times 10^{52}$ erg/s~\cite{Raffelt:1996wa}:
\be
\label{eq:SNbound}
\mathcal{L}_J \simeq Q_J \left(\frac{4\pi}{3} \, R_c^3\right) \leq \mathcal{L}_\nu~\Rightarrow~
f \gtrsim \text{MeV} \times \left( m_J / \text{keV} \right)
~, 
\ee
where $R_c \simeq 10$ km is the core radius and we have taken $m_\nu = 0.1 \eV$ 
to maximize 
the energy loss.
Our estimate is in good agreement with the dedicated analyses performed in
Refs.~\cite{Choi:1989hi,Kachelriess:2000qc,Farzan:2002wx,Heurtier:2016otg}. The lower bound on $f$ in Eq.~(\ref{eq:SNbound})
 is orders of magnitude below the thermally-favored values in
Fig.~\ref{fig:fFO}. Other relevant processes involving neutrinos include
neutrino annihilation into pairs of majorons, i.e., $\nu \nu \to J J$. 
However, compared to single majoron production, this rate is suppressed by an
additional factor of $(m_\nu / f)^2 \ll 1$. 
Finally, we note that right-handed neutrinos with a mass of $M_N\sim 200\MeV$ can help 
restart stalled shock-fronts and facilitate supernovae explosions~\cite{Fuller:2009zz}. This 
is precisely in the cosmologically motivated region in Fig.~\ref{fig:fFO} for $M_N \sim f$.

As discussed in Sec.~\ref{sec:scalar}, interactions of $J$ with SM leptons also arise from loops of intermediate sterile and active neutrinos. For instance, loop-induced electron Yukawas are parametrically of size $m_\nu m_e / 16 \pi^2 v^2 \sim 10^{-20}$. These are well below the upper bounds derived from anomalous cooling of red giants and horizontal branch stars in Ref.~\cite{Hardy:2016kme}.

\subsection{Direct Searches}
\label{sec:direct}

Another avenue in exploring these models consists of direct searches for the light HS mediators ($J,S, N$) and/or DM ($\x$). As discussed in detail in Ref.~\cite{Garcia-Cely:2017oco}, limits on majoron-SM couplings are obtained from searches for flavor-violating processes, such as neutrinoless double beta decay, $K \to \pi J$, and $\mu \to e J$, which constrain $m_\nu / f \lesssim 10^{-5} - 10^{-2}$, corresponding to $f \gtrsim 10 \eV - 10 \keV$~\cite{Pilaftsis:1993af,Feng:1997tn,Hirsch:2009ee,Rodejohann:2011mu,Blum:2018ljv}. Furthermore, for sterile neutrinos near the $U(1)_L$-breaking scale, $f \sim 100 \MeV$, measurements of meson decays, such as $\pi, K \to \ell \, \nu$ are also potentially relevant and are sensitive to active-sterile mixing at the level of $m_\nu / M_N \lesssim \text{few} \times 10^{-9} - 10^{-8}$. See Ref.~\cite{deGouvea:2015euy} for a comprehensive review of such searches. While these limits are not sensitive to the natural parameter space of these models, they exclude non-trivial forms of the active-sterile mixing matrix, $R$ (see Eq.~(\ref{eq:V2})), that lead to enhanced mixing in the neutrino sector.

Recent years have seen an increased focus on new experimental technologies to explore the sub-GeV DM frontier~\cite{Battaglieri:2017aum}. Of particular interest in this work are futuristic detectors proposed to detect elastic recoils of nucleons or electrons from DM as light as $\sim \order{\text{keV}}$, corresponding to $\sim \order{\text{meV}}$ energy depositions~\cite{Hochberg:2015pha,Hochberg:2015fth,Schutz:2016tid,Knapen:2016cue}. In this section, we investigate the potential sensitivity of these experiments to the classes of models discussed throughout this work. 

The strength of $\x-\text{SM}$ elastic scattering is controlled by the size of the $S-h$ and $J-h$ mixing angles, $\alpha$ and $\beta$, respectively (defined in Appendix~\ref{sec:scalar_masses}). For the cosmologically-favored parameter space in Fig.~\ref{fig:fFO}, Eq.~(\ref{eq:tachyon2}) suggests that for $m_\x \sim (1- 100) \keV$, $\beta \lesssim 10^{-16} - 10^{-12}$ is needed to avoid tachyonic states in the HS scalar spectrum. The prospects for such couplings to yield detectable rates is minuscule, and hence, 
$J$-mediated interactions with charged SM fermions are negligible within the context of direct detection experiments. In contrast, the $S-h$ mixing angle, $\alpha$, is not as constrained, so we focus on $S$-mediated interactions. 
The Yukawa coupling of the SM fermions to $S$ is given by
\be
\mathscr{L} \supset - \, \frac{\alpha \, m_f}{v} \, S \, \bar{f} f
~.
\ee
This can be matched onto a low-energy theory involving nucleons ($n$) and pions ($\pi^\pm$)~\cite{Gunion:1989we},
\be
\mathscr{L} \simeq - \, \frac{\alpha}{v} ~ S ~ \left[ \frac{4}{29} \, m_n \, \bar{n} n  + \frac{2}{9} \, \left( m_S^2 + \frac{11}{2} \, m_\pi^2 \right) \pi^+ \pi^- \right]
~.
\ee
For $m_S \sim 10 - 100 \MeV$, the most stringent limits on $\alpha$ arise from considerations of anomalous cooling of SN1987A from the emission of $S$~\cite{Ishizuka:1989ts,Krnjaic:2015mbs}. Such production is strongly suppressed when $m_S \gtrsim 200 \MeV$, and we instead bound $\alpha$ by demanding that the processes $S \pi \leftrightarrow \gamma \pi$, $S p \leftrightarrow \gamma p$, $S e \leftrightarrow \gamma e$, and $S \leftrightarrow e^+ e^-$ do not prematurely equilibrate the HS and SM at temperatures below the QCD phase transition. For reheat temperatures at the level of $T_\text{RH} \sim 5 \MeV$ and $m_S \sim 10 - 100 \MeV$, equilibration through $S-h$ mixing does not occur for $\alpha \lesssim 10^{-5} - 10^{-3}$, respectively. In this mass range, considerations of SN1987A constrain mixing angles larger than $\alpha \sim 10^{-6}$. If $\alpha$ is set to its maximally allowed value and $f$ is fixed to the thermal line in Fig.~\ref{fig:fFO}, we find that the DM-nucleon elastic scattering rate is well below the irreducible neutrino background, $\sigma_p \ll 10^{-50} \text{ cm}^2$, while the electron scattering rate is many orders of magnitude below the sensitivities of futuristic proposed technologies~\cite{Battaglieri:2017aum}.

We now consider variations upon these minimal models. We will first propose a modification in which the scalar mediator $S$ is lighter than $\x$ and the scale $f$ and possesses additional couplings to the SM. As in Ref.~\cite{Knapen:2017xzo}, we assume that $S$ also couples directly to SM QCD, through an interaction of the form
\be
\label{eq:GG}
\mathscr{L} \sim \frac{1}{\Lambda} ~ S ~ G^{a}_{\mu \nu} G^{a \mu \nu}
~,
\ee
where $\Lambda$ is the cutoff of the effective theory. This interaction could be generated, for instance, from direct couplings to a vector-like generation of heavy quarks. As before, this can be mapped onto a theory involving nucleons and pions at low energies. Parametrically, this is of the form
\be
\mathscr{L} \sim y_n ~ S ~ \bar{n} n + \frac{y_n}{m_n} ~ S ~ \partial_\mu \pi^\dagger \partial^\mu \pi
~.
\ee
As shown explicitly in Ref.~\cite{Knapen:2017xzo}, these couplings can lead to detectable
rates in proposed low-threshold detectors for $m_\x \sim m_S \sim 100 \keV$,
without conflicting with cosmological, astrophysical, or terrestrial
constraints. In order to enlarge the viable parameter space, we propose a
slight modification of the model in Ref.~\cite{Knapen:2017xzo}, which we now
outline. 

Compared to canonical WIMPs, physics at temperatures much greater than  $\sim \text{MeV}$ is not directly important for models of sub-MeV thermal relics. In light of this, we will consider a low reheat temperature of the universe following inflation, $T_\text{RH}$. 
The requirement of radiation domination during BBN implies that $T_\text{RH} \gtrsim \text{few} \MeV$~\cite{Hannestad:2004px,deSalas:2015glj}. We will take
\be
T_\text{RH} \sim 5 - 10 \MeV
\ee
for concreteness. This is also motivated in models involving gravitinos and/or moduli~\cite{Pradler:2006hh,Kohri:2005wn,Moroi:1999zb}. We now ask: what are the maximum allowed values of the nucleon coupling, $y_n$, such that the DM and visible sectors do not equilibrate before neutrino-photon decoupling? The decays and inverse-decays, $J \leftrightarrow \nu \nu$, are still assumed to equilibrate the two sectors below a few MeV. We find that processes involving protons, $p$, and pions, $\pi$, such as $S p \leftrightarrow \gamma p $ and $S \pi \leftrightarrow \gamma \pi $ do not equilibrate the two sectors before neutrino-photon decoupling provided that $y_n \lesssim 10^{-5} - 10^{-3}$, where the lower (upper) part of the range corresponds to $T_\text{RH} \sim 10 \, (5) \MeV$, respectively. By closing a loop of charged nucleons or pions, these couplings also generate an interaction with photons, which (modulo tuning) is naturally of size
\be
\label{eq:photoncoupling}
\mathscr{L} \sim \frac{\alpha_\text{em} \, y_n}{4 \pi \, m_n} ~ S ~ F_{\mu \nu} ~ F^{\mu \nu}
~.
\ee
We demand that the processes $S \leftrightarrow \gamma \gamma$ also does not prematurely equilibrate the DM and visible sectors. This leads to the additional upper bound $y_n \lesssim 10^{-4} \, (m_S  / 100 \keV)^{-1/2}$. Hence, in order for equilibration to occur below the temperature of neutrino-photon decoupling, we will conservatively require that $y_n \lesssim 10^{-5}$.

An exhaustive study of the constraints on DM-nucleon couplings in the context of MeV-scale particles has recently been presented in Ref.~\cite{Knapen:2017xzo}.
Here, we summarize the most relevant bounds.
Considerations of cooling of horizontal branch stars constrain $y_n \ll 10^{-10}$. However, this limit rapidly diminishes for $m_S \gtrsim 100 \keV$. For masses above $\sim 200 \keV$, the dominant constraints are from measurements of the meson decays, $K \to \pi S$, leading to $y_n \lesssim 10^{-5}$. For $y_n \gtrsim 10^{-7}$, $S$ is produced but trapped in supernova,
and bounds from anomalous cooling are evaded. 
Therefore, limits from meson decays and stellar/supernovae cooling restrict the nucleon coupling to be in the range
\be
10^{-7} \lesssim y_n \lesssim 10^{-5}
\quad \text{(viable range)}
~,
\ee
for $m_S \gtrsim 100 \keV$. As argued above, for couplings of this size, DM-SM equilibration in the early universe 
is still driven by the neutrino-majoron coupling, as in our minimal scenario of Sec.~\ref{sec:equil}.
The DM-proton elastic scattering cross section is roughly
\be
\sigma (\x p \to \x p) \sim \frac{y_n^2}{4 \pi} ~ \frac{m_\x^4}{f^2 \, m_S^4}
~.
\ee
For $m_S \gtrsim 100 \keV$, and taking $y_n \sim 10^{-6}$, we have
\be
\sigma (\x p \to \x p) \sim 10^{-40} \text{ cm}^2 \times \left( \frac{m_\x}{m_S} \right)^4 ~ \left( \frac{m_\x}{200 \keV} \right)^{-1}
\ee
where we have fixed $f$ to the thermally-favored value in Fig.~\ref{fig:fFO}. Proposed experiments, such as superfluid helium targets, are projected to be sensitive to cross sections as small as $\sigma_\text{DD}^p \sim 10^{-42} \text{ cm}^2$ in this mass range~\cite{Battaglieri:2017aum}.

\section{Summary and Conclusions}
\label{sec:conclusion}

In recent years, there has been growing interest in exploring new cosmological paradigms and modes of detection for particle dark matter in the $\text{keV}-\text{GeV}$ mass range. For such light masses, dark matter that is of a thermal origin is strongly constrained from a plethora of cosmological and astrophysical considerations, including nucleosynthesis, the cosmic microwave background, structure formation, and stellar cooling. In particular, sub-MeV thermal relics that were in equilibrium with the Standard Model bath at temperatures below an $\text{MeV}$ necessarily contribute to deviations in the expansion rate of the universe at the time of nucleosynthesis and/or recombination relative to the standard cosmology. As a result, models of sub-MeV thermal dark matter are usually thought to be either excluded or require involved model-building to evade these constraints. 

We have focused on a class of models that naturally evade such claims. For instance, if a cold hidden sector equilibrates with the Standard Model after neutrino-photon decoupling, deviations in the expansion rate of the universe are strongly suppressed, alleviating the corresponding bounds from measurements of the effective number of neutrino species. Although this statement applies to dark matter that equilibrates either with neutrinos or photons, we have focused on interactions with the Standard Model neutrino sector. This is motivated, in part, by the fact that constraints derived from stellar cooling are much stronger for new light forces that couple directly to electromagnetism.

We studied concrete realizations of the above scenario where the dark sector masses and interactions, as well as the observed neutrino masses and mixing angles, are generated at a single scale corresponding to the spontaneous breaking of lepton number in the Standard Model.
The pseudo-Goldstone boson associated with this breaking is the majoron, which is the mediator responsible for equilibrating the dark matter and Standard Model sectors in the early universe. These models independently motivate the sub-MeV scale; demanding that thermal dark matter freezes out with an adequate abundance implies that its mass is parametrically related to the Planck mass, the temperature at matter-radiation equality, and the measured neutrino masses by $\mdm \sim (m_\text{Pl} / T^\text{MRE})^{1/4} \, m_\nu \sim \text{MeV}$. Along with considerations of structure formation, this restricts the viable mass range to $\mdm \sim 10 \keV - \text{MeV}$ and the majoron-neutrino interaction strength to be at the $10^{-10}$ to $10^{-9}$ level. 

Despite the suppressed size of such interactions, this class of models will be decisively tested in the near future. For instance, thermal relics that relativistically equilibrate with any Standard Model species after neutrino-photon decoupling lead to an irreducible deviation in the effective number of neutrino species above the projected sensitivity of future CMB-S4 experiments. 
Improved measurements of the small- and large-scale structure of the universe will also probe 
these models,
potentially testing most of the remaining parameter space.
Furthermore, 
it is possible to introduce a large coupling of the majoron to nucleons which preserves the viability of the cosmology provided that the reheat temperature of the universe is small ($\sim 10 \MeV$). In this case, dark matter detection is possible at recently proposed low-threshold direct detection experiments aimed at exploring the sub-GeV dark matter frontier.

\bigskip
\bigskip
 
\begin{acknowledgments}
  We thank Lawrence Hall, Keisuke Harigaya, Simon Knapen, Gustavo Marques-Tavares, David Morrissey, David McKeen, and Maxim Pospelov for valuable discussions. AB and NB are supported by the U.S. Department of Energy under Contract No. DE-AC02-76SF00515. Part of this work was completed at the Kavli Institute for Theoretical Physics, which is supported in part by the National Science Foundation under Grant No. NSF PHY-1748958. 
NB thanks TRIUMF for hospitality during the completion of this work. 
\end{acknowledgments}

\appendix
\section{Model Details\label{sec:majoron_int}}

\subsection{Fermion Masses and Interactions}
\label{sec:app1}

In this Appendix we summarize our conventions and present majoron and 
neutrino interactions in the mass basis. Our conventions mostly follow those of 
Refs.~\cite{Pilaftsis:1993af,Garcia-Cely:2017oco}. First, we obtain a useful parametrization of the 
neutrino mixing matrix, $V$ (see Eq.~(\ref{eq:V1})), and the associated interactions in the seesaw limit ($m_D/M_N \ll 1$) where 
the active neutrino mass matrix reduces to
\be
\label{eq:seesaw1}
M_\nu = - m_D ~ M_N^{-1} ~ m_D^T
~
\ee
after integrating out the right-handed neutrinos.
Diagonalizing $M_\nu$ gives the $3 \times 3$ matrix, 
\be
d_\ell = \text{diag}(m_1, m_2, m_3),
\ee
where $m_{1,2,3}$ are the masses of the SM neutrinos. In general, the phases of $N$ can be chosen such that $M_N$ is purely diagonal, $M_N = d_h$. 
In the seesaw limit, 
\be
d_h \simeq \text{diag}(m_4, m_5, m_6),
\ee
where $m_{4,5,6}$ ($\gg m_{1,2,3}$) are the masses of the sterile neutrinos. 
The Dirac matrix, $m_D$, can be generally decomposed in the Casas-Ibarra form~\cite{Casas:2001sr}
\be
\label{eq:CIparam}
m_D = i \, U \, \sqrt{d_\ell} \, R^T \, \sqrt{d_h}
~,
\ee
where $R$ is a complex orthogonal $3 \times 3$ matrix that parametrizes mixing between the active-sterile species. For simplicity, we will set $R = \mathbb{1}$. As noted in Ref.~\cite{Casas:2001sr}, this choice of $R$ corresponds to the special case in which $y_\nu$ and $M_N$ are simultaneously diagonalizable, while the charged lepton sector is not. This corresponds to a model in which all of the lepton flavor violation originates from the charged lepton sector. $U$ is the standard PMNS matrix, whose entries are fixed by the known neutrino mixing angles. Eq.~(\ref{eq:CIparam}) can be proved by the following argument. We define the unitary PMNS matrix such that it diagonalizes $M_\nu$,
\be
\label{eq:PMNS}
U^\dagger ~ M_\nu ~ U^* = d_\ell
~.
\ee
Using Eq.~(\ref{eq:seesaw1}), we can rewrite Eq.~(\ref{eq:PMNS}) as
\begin{align}
& - U^\dagger \, m_D \, d_h^{-1} \, m_D^T \, U^* = d_\ell
\implies \left( \sqrt{d_\ell^{-1}} \, U^\dagger \, m_D \, \sqrt{d_h^{-1}} \right) \times \left(  \sqrt{d_\ell^{-1}} \, U^\dagger \, m_D \, \sqrt{d_h^{-1}} \right)^T = -1
~,
\end{align}
which implies that
\be 
\sqrt{d_\ell^{-1}} \, U^\dagger \, m_D \, \sqrt{d_h^{-1}} = i \, R^T
~,
\ee
where $R$ is any complex matrix such that $R \, R^T = \mathbb{1}$. Solving for $m_D$ gives Eq.~(\ref{eq:CIparam}). 
As stated in Ref.~\cite{Casas:2001sr}, continuous forms of $R$ (not including reflections) can be parametrized in terms of three complex angles.
In the seesaw limit, $V$ takes the form
\be
\label{eq:V2}
V \simeq \begin{pmatrix} U^* & - i U^* \sqrt{d_\ell} \, R^\dagger \sqrt{d_h^{-1}} \\ - i \sqrt{d_h^{-1}} \, R \sqrt{d_\ell} & \mathbb{1} \end{pmatrix}
~.
\ee
It is straightforward to check that Eqs.~(\ref{eq:V1}) and (\ref{eq:V2}) hold to leading order in $d_\ell / d_h$.
The off-diagonal entries in Eq.~(\ref{eq:V2}) parametrize the active-sterile neutrino mixing.

Electroweak- and $U(1)_L$-breaking leads to mixing amongst the neutrino states. We now switch to four-component notation and denote the Majorana neutrino mass eigenstates as $n_i$, $i = 1,2,\dots,6$, with mass $m_i$, such that $n_{1,2,3}$ are SM-like, and $n_{4,5,6}$ are sterile-like. We parametrize the couplings of these states to the scalar sector as
\be
\mathscr{L} \supset  
  J ~ \bar{n}_i \left(  \lambda_{J s}^{(ij)} + i \gamma^5 ~ \lambda_{J p}^{(ij)}\right) n_j 
+ S ~ \bar{n}_i \left(  \lambda_{S s}^{(ij)} + i \gamma^5 ~ \lambda_{S p}^{(ij)} \right) n_j 
+ h ~ \bar{n}_i \left(  \lambda_{h s}^{(ij)} + i \gamma^5 ~ \lambda_{h p}^{(ij)} \right) n_j
~,
\label{eq:majoron_neutrino_int}
\ee
where $h$ is the physical SM Higgs field. As shown in Ref.~\cite{Pilaftsis:1993af}, the effective couplings are
\begin{align}
\lambda_{J s}^{(ij)} \equiv \frac{1}{2 f} \left( m_j - m_i \right) \text{Im} \, C_{ij}
\quad &,\quad
\lambda_{J p}^{(ij)} \equiv \frac{1}{2 f} \left( m_i + m_j \right) \Big( \frac{1}{2} \, \delta_{ij} - \text{Re} \, C_{ij} \Big) 
\nl
\lambda_{S s}^{(ij)} \equiv \frac{1}{2 f}  \left( m_i + m_j \right) \Big( \text{Re} \, C_{ij} - \frac{1}{2} \delta_{ij} \Big)
\quad &, \quad
\lambda_{S p}^{(ij)} \equiv \frac{1}{2 f} \left( m_j - m_i \right) \text{Im} \, C_{ij}
\nl
\lambda_{h s}^{(ij)} \equiv - \, \frac{1}{2 v}  \left( m_i + m_j \right) \text{Re} \, C_{ij} 
\quad &, \quad
\lambda_{h p}^{(ij)} \equiv \frac{1}{2 v} \left( m_i - m_j \right) \text{Im} \, C_{ij}
~,
\label{eq:Cij}
\end{align}
where following Ref.~\cite{Pilaftsis:1993af}, we define
\be
C_{ij} \equiv \sum\limits_{k=1}^3 V_{ki} V_{kj}^*
~.
\ee
In general, there may be other contributions to the masses of the sterile neutrinos. In this case, the mass parameters $m_{4,5,6}$ written in Eq.~(\ref{eq:Cij}) are interpreted as the piece given by the scale $f$, i.e., $\sim f \times \partial m / \partial f$. The interactions of the electroweak gauge bosons with the neutrinos are given by
\be
\mathscr{L} \supset 
Z_\mu ~ \bar{n}_i \gamma^\mu \left( i \, g_{Z v}^{(ij)} + g_{Z a}^{(ij)} \gamma^5 \right) n_j 
+ \left[ g_W^{(ij)} ~ W_\mu^- ~  \bar{\ell}_i \gamma^\mu (1 - \gamma^5) n_j + \text{h.c.} \right]
~,
\label{eq:neutrino_gauge_int}
\ee
where the couplings are defined as
\begin{align}
g_{Z v}^{(ij)} &\equiv - \, \frac{g_2}{4 c_w} ~ \text{Im} \, C_{ij}
\nl
g_{Z a}^{(ij)} &\equiv \frac{g_2}{4 c_w} ~ \text{Re} \, C_{ij}
\nl
g_W^{(ij)} &\equiv - \, \frac{g_2}{2 \sqrt{2}} ~ B_{i j}
~,
\end{align}
and 
\be
B_{i j} \equiv \sum\limits_{k=1}^3 \delta_{i k} V_{k j}^*
~.
\ee
Note that $V$ is a $6 \times 6$ matrix, but the sum above is only over the first three indices, i.e., the active-like states. 
Using the 
seesaw expression for $V$ in Eq.~(\ref{eq:V2}), $C$ and $B$ can be written as
\begin{align}
C &\simeq \begin{pmatrix} \mathbb{1} & i \sqrt{d_\ell} \, R^T \sqrt{d_h^{-1}} \\ - i \sqrt{d_h^{-1}} \, R^* \sqrt{d_\ell} & 0 \end{pmatrix}
\nl
\nl
B &\simeq \begin{pmatrix} U && i U \sqrt{d_\ell} \, R^T \sqrt{d_h^{-1}} \\ 0 && 0 \end{pmatrix}
~.
\end{align}

\subsection{Scalar Masses}
\label{sec:scalar_masses}
The most general renormalizable potential with soft $U(1)_L$-breaking is given by 
\be
V = - \mu_H^2 \, |H|^2 + \lambda_H \, |H|^4 - \mu_\sigma^2 \, |\sigma|^2 + \lambda_\sigma \, |\sigma|^4  + \lambda_{\sigma H} \, |\sigma|^2 |H|^2
- \left( \mu_\sigma^{\prime \, 2} \, \sigma^2 + a_\sigma \, \sigma \, |H|^2 \right).
\ee
We fix the phase of $\sigma$ such that its vev is real; the phases of the Yukawa couplings $\lambda_\x$ and $y_\nu$ defined in Eqs.~(\ref{eq:Lag1}) and (\ref{eq:ds_yukawa}) 
are fixed such that the resulting fermion mass contributions are real. This leaves a single 
physical phase in the model shared between the parameters $\mu_\sigma^\prime$ and $a_\sigma$. 
The potential minimization conditions, $\partial V / \partial v = \partial V / \partial f = \partial V/\partial J = 0$, 
can be solved for $\mu_{H, \sigma}^2$ and the imaginary parts of the soft terms,
\begin{align}
&\mu_H^2 = \lambda_H \, v^2 + \left( \frac{1}{2} \, \lambda_{\sigma H} - \frac{\sqrt{2} ~ \text{Re} \, a_\sigma }{f} \right) \, f^2
\nl
&\mu_\sigma^2 = \lambda_\sigma \, f^2 + \left( \frac{1}{2} \, \lambda_{\sigma H} - \frac{\text{Re} \, a_\sigma }{\sqrt{2} ~  f} \right) \, v^2 - 2 \, \text{Re} \, \mu_\sigma^{\prime \, 2}
\nl
&\text{Im} \, \mu_\sigma^{\prime \, 2} = - \, \frac{\text{Im} \, a_\sigma \, v^2}{2 \sqrt{2} \, f}
~.
\end{align}
Imposing these conditions, the scalar mass matrix in the $(J, S, h)$ basis simplifies to
\be
M_\p^2 = 
\begin{pmatrix}
m_J^2 & - \, \frac{\text{Im} \, a_\sigma}{\sqrt{2} \, f} \, v^2 & \sqrt{2} \, \text{Im} \, a_\sigma \, v 
\\
- \, \frac{\text{Im} \, a_\sigma}{\sqrt{2} \, f} \, v^2 & m_S^2 & \left(  \lambda_{\sigma H} - \frac{\sqrt{2} \, \text{Re} \, a_\sigma}{f} \right) v \, f
\\
\sqrt{2} \, \text{Im} \, a_\sigma \, v & \left(  \lambda_{\sigma H} - \frac{\sqrt{2} \, \text{Re} \, a_\sigma}{f} \right) v \, f & m_h^2 
\end{pmatrix}
~,
\ee
where the diagonal entries correspond to the masses of the unmixed fields:
\begin{align}
m_J^2 & = 4 \Re \musigp + \frac{\Re \asig v^2}{\sqrt{2}f }, \\
m_S^2 & = 2 \lsig f^2 + \frac{\Re \asig v^2}{\sqrt{2} f},\\
m_h^2& = 2\lambda_h v^2
~.
\end{align}
The mass matrix $M_\p^2$ is diagonalized in the mass eigenstate basis, given by $\p_{1,2,3}$.
In the limit of small mixing the flavor eigenstates are related to $\p_{1,2,3}$ via 
\be
\begin{pmatrix}
  J \\ 
  S \\
  h 
\end{pmatrix}
= 
\begin{pmatrix}
   1 & -\gamma  & \beta  \\
    \gamma  & 1 & -\alpha  \\
     -\beta  & \alpha  & 1 \\
\end{pmatrix}
\begin{pmatrix}
  \varphi_1 \\ 
  \varphi_2 \\
  \varphi_3 
\end{pmatrix},
\label{eq:mixingmatrix}
\ee
where the small angles $\alpha$, $\beta$, and $\gamma$ are defined by
\begin{align}
  \alpha & = - \frac{\left(\lsh -\frac{\sqrt{2}\Re\asig}{f}\right)v f}{(m_h^2 - m_S^2)} \, ,\\
  \beta & = \frac{\sqrt{2}\Im\asig v}{(m_h^2 - m_J^2)} \, , \\
  \gamma & = \frac{\Im\asig v^2}{\sqrt{2}f(m_S^2-m_J^2)} \, .
\end{align}

Large mixing in the scalar sector can lead to tachyonic masses.
The most stringent constraint is obtained in the $S-J$ sector 
(since $J$ is the lightest state and mixing with $h$ is 
suppressed by the large Higgs mass). Requiring that the 
$S-J$ eigenstates have positive masses bounds the mixing as
\be
\gamma < \frac{m_{S} m_{J}}{m_{S}^2-m_{J}^2} \, .
\label{eq:tachyon}
\ee
This constraint can also be translated into a bound on $\beta$ 
\be
\beta < \frac{2 f m_J m_S}{v m_h^2} \, ,
\label{eq:tachyon2}
\ee
which limits the size of tree-level interactions of the majoron with charged SM fermions (see Appendix~\ref{sec:chargedfermi}).


\subsection{Scale of Lepton-Number Breaking and Planckian Effects}
\label{sec:planck}
In this Appendix, we briefly comment on two theoretical aspects of the majoron construction described above.
We have introduced two new energy scales associated with spontaneous and explicit $U(1)_L$-breaking, $f$ and $m_J$, respectively, 
with $f\gg m_J$. As we saw in Sec.~\ref{sec:cosmo}, considerations of DM-SM equilibration require $f$ to be much smaller than the electroweak scale, i.e., $f \ll v \simeq 246 \GeV$.

The first issue associated with these new energy scales is the radiative stability of $f$. Quantum corrections 
will generically shift the mass term (and the resulting vev) of the $U(1)_L$-breaking field, $\sigma$, to the UV cutoff of the theory, i.e., 
$\Lambda \gg v$. As with the SM Higgs hierarchy problem, supersymmetry can be used to regulate the sensitivity to UV physics.
If the HS (including $\sigma$) is sequestered from the supersymmetry-breaking sector, a naturally small $f$ can be 
radiatively induced through interactions with the SM via the right-handed (s)neutrino~\cite{Chacko:2003dt}. 
However, a small supersymmetry-breaking scale in the HS also implies the presence of new light degrees of freedom 
(e.g. the superpartners of $\x$ and $\p$) that can play an important role in cosmology. A detailed investigation of this scenario is beyond the scope of this work.

The second puzzling feature of the majoron construction is the origin of the scale $m_J$.
If $U(1)_L$ was an exact symmetry (at least classically), the majoron would be massless, so $m_J > 0$ requires an 
explicit breaking of $U(1)_L$. While the hierarchy $m_J\ll f$ is protected by the fact that $J$ is a pNGB,
it is interesting to ask why $m_J < f$ in the first place if they are completely unrelated.
Global symmetries are expected to be absent in 
theories of quantum gravity. A simplified argument is that a scattering process 
with a global charge in the initial state can destroy the charge in an intermediate black hole state. 
The black hole cannot carry 
global charge, so it decays democratically via Hawking radiation~\cite{Kamionkowski:1992ax,Holman:1992va}. 
This means that the low-energy effective field theories should have 
Planck-scale violations of global symmetries. This is a well-known 
problem in axion models with a global Peccei-Quinn $U(1)$~\cite{Holman:1992us,Kamionkowski:1992mf,Barr:1992qq}. 
Thus $U(1)_L$-breaking effects should also appear 
in the low energy description~\cite{Akhmedov:1992hi,Rothstein:1992rh}.

If the Planck-scale effects are unsuppressed, then one expects 
mass terms $\sim \Mpl^2 \sigma^2$ to appear, which would remove 
any pNGB from the spectrum. Thus, if we want a light majoron, 
Planck effects should enter through marginal or irrelevant 
operators $\sim 1/\Mpl^n$, $n\geq 0$. The standard way to ensure this is 
to engineer $U(1)_L$ to be an accidental symmetry, 
i.e., one that is a consequence of gauge charge assignments as in Ref.~\cite{Rothstein:1992rh}. 
This can be accomplished, e.g., using a gauged $U(1)_{B-L}$ with an 
additional scalar field $\varphi$, such that the leading $U(1)_L$-breaking term is
\be
\mathscr{L} \supset \frac{\sigma^{n_1} \varphi^{n_2}}{\Mpl^{n_1 + n_2 - 4}} + \hc \, ,
\ee
where the integer powers $n_{1,2}$ are determined by the charge assignments $Q_{B-L}[\sigma]$ and $Q_{B-L}[\varphi]$.\footnote{$B-L$ is only anomaly-free \emph{after} including 3 RH neutrinos~\cite{Langacker:2008yv}. When we include DM, it must also 
be charged under $B-L$ (since it couples to $\sigma$), so the anomaly must be 
canceled again by some additional states.}
For example if $Q_{B-L}[\sigma] = -2$ and $Q_{B-L}[\varphi] = \{1/2,\, 4/3,\, 3,\, 8\}$, the 
lowest-dimensional $L$-breaking operators are dimension-five~\cite{Rothstein:1992rh}:
\be
\frac{1}{\Mpl}\left\{
 \sigma \varphi^4,~ \sigma^2 \varphi^3,~ \sigma^3 \varphi^2,~ \sigma^4 \varphi  
\right\}.
\ee
When $\varphi$ gets a vev, these operators can be mapped onto the $L$-breaking terms in the 
$\sigma$ potential in Eq.~(\ref{eq:full_pot}). 
Note that for a given charge assignment with this minimal field content, only one of the 
potential terms is generated at dimension five. This means that it is a reasonable approximation to 
turn them on one at a time in this minimal framework. 

Is there a mass-scale that is singled out by the Planck-suppressed operators? 
The answer depends on what the natural scale for spontaneous $B-L$ breaking is. 
At the very least, one needs to account for existing bounds on the $B-L$ gauge boson, 
a type of $Z^\prime$ which has been 
extensively studied, see, e.g. Refs~\cite{Carena:2004xs,Langacker:2008yv,Basso:2008iv}.
The LHC constrains $m_{Z'}/g_{Z'} > 6-100\;\TeV$ through 
dilepton resonance searches and bounds on 
four-fermion contact interactions~\cite{Aaboud:2017buh,Sirunyan:2018exx}. 
Letting $\langle \varphi \rangle=  v_{B-L}/\sqrt{2}$ 
and $m_{Z'} = q_\varphi g_{Z'} v_{B-L}$, the above experimental bound implies
\be
q_\varphi v_{B-L} \gtrsim 6-100\;\TeV,
\ee
where the range depends on the mass of the $Z^\prime$.
In the minimal scenario with $q_\varphi = Q_{B-L}[\varphi]= 4/3$, the $\musigp \sigma^2$ term is 
generated from a dimension-five operator
\be
\frac{1}{\Mpl} \sigma^2 \varphi^3 \rightarrow \left(\frac{\vbl^3}{2\sqrt{2}\Mpl}\right)\sigma^2.
\ee
The experimental bound then suggests a very rough lower limit on the majoron mass
\be
m_J^2 \sim \musigp \sim \frac{\vbl^3}{\Mpl} \gtrsim (100\;\keV)^2.
\ee
This bound can be much weaker if the mass is generated by an operator with a 
higher dimension or if its Wilson coefficient is not 
$\mathcal{O}(1)$. It can be larger if $U(1)_L$ is explicitly broken at a scale $\Lambda<\Mpl$, e.g., the GUT scale.
Thus, the natural size for the majoron mass (\emph{if $B-L$ is broken near the weak-scale and the scale of explicit breaking is $\Mpl$}) is 
near the keV-scale under the above assumptions. This was also noted in Ref.~\cite{Akhmedov:1992hi}.
While this link is tenuous at best, it is reassuring that an internally consistent picture for the scales $f$ and $m_J$ seems 
attainable.

\subsection{Interactions with Charged Fermions}
\label{sec:chargedfermi}

  The mixing of the dark sector scalars with the SM Higgs gives 
  rise to $S$ and $J$ coupling to SM fermions.
  These interactions can be summarized by 
\be
\mathscr{L} \supset \frac{m_f}{v}\left[\beta J - \alpha S - (1 - \alpha^2/2 -\beta^2/2)h\right]\bar f f \, ,
\ee
where we approximated $\varphi_1 \simeq J$, $\varphi_2 \simeq S$ and $\varphi_3 \simeq h$. Note that 
the interactions of the 125 GeV Higgs-like state are suppressed relative to the SM expectation 
by an effective mixing 
\be
\cos \theta_{\mathrm{eff}} \simeq 1 - \alpha^2/2 -\beta^2/2 \, .
\ee

The strongest constraints on the scalar potential parameters come from
rare meson and invisible Higgs decays. These were recently analyzed 
in Ref.~\cite{Krnjaic:2015mbs} in the context of Higgs-portal coupled dark sectors. 
A detailed discussion of flavor physics constraints is presented in Ref.~\cite{Dolan:2014ska}.
An invisibly-decaying 
light scalar, $\varphi$, that mixes with the SM Higgs contributes to 
the invisible decay modes $B^\pm \rightarrow K^\pm \varphi$ and 
$K^\pm \rightarrow \pi^\pm \varphi$, whenever kinematically allowed. 
We are interested in $J$ and $S$ that are much lighter
than $m_B - m_K$ and $m_K - m_\pi$, so that the 
observed limits on these rare decay modes constrain the 
effective mixing
\begin{align}
  \sin^2 \theta_{\mathrm{eff}}  < 9\times 10^{-6} & \;\;(B^\pm \rightarrow K^\pm + \mathrm{inv.}) \\ 
  \sin^2 \theta_{\mathrm{eff}}  < 3 \times 10^{-8} & \;\;(K^\pm \rightarrow \pi^\pm + \mathrm{inv.}). 
\end{align}

Measurements of the Higgs properties at the LHC also constrain the parameters of the scalar potential.
For example, since $\x$, $N$, $S$, and $J$ are much lighter than $h$, there are new invisible decay modes.
The invisible branching fraction of the Higgs is constrained to be less than $0.23$ at 95\% confidence level~\cite{Aad:2015pla,Khachatryan:2016whc}, 
leading to the bound 
\be
\left(\frac{m_\x}{2f}\right)^2 \sin^2 \theta_{\mathrm{eff}} + \lsh^2 \left(\frac{v^2}{2m_h^2}\right) 
+ \sum_{i,j} \left[\left(\lambda_{h s}^{(ij)}\right)^2 + \left(\lambda_{h p}^{(ij)}\right)^2\right]
< 2\times 10^{-4},
\ee
where the terms correspond to $h\rightarrow \x\x$, $h\rightarrow SS,\; JJ$, and 
$h \rightarrow n_i n_j$, respectively.

Interactions of $S$ and $J$ with the charged SM fermions 
are also generated by loops of neutrinos via couplings in Eqs.~(\ref{eq:majoron_neutrino_int}) and (\ref{eq:neutrino_gauge_int})
~\cite{Pilaftsis:1993af,Garcia-Cely:2017oco}.
Their characteristic size (see Eq.~(\ref{eq:loopmixing})) corresponds to a 
tiny effective mixing of $\sim 10^{-15}$.
Thus, even with the stringent constraints on the mixing angles, the tree-level 
interactions of $S$ and $J$ with charged SM fermions can be much larger than those 
induced by loops. 
\bibliography{majoron}

\end{document}